\documentclass[a4paper,12pt,preprint,amsmath,showpacs,nofootinbib,superscriptaddress]{revtex4-1}
\usepackage{amsmath}
\usepackage{graphicx}
\usepackage{bm,braket}
\usepackage{multirow}
\usepackage{color}
\usepackage{slashed}
\usepackage{extarrows}

\newcommand{\als}{\alpha_s}

\newcommand{\nn}{\nonumber}
\newcommand{\ff}{f\hspace{-0.4em}f}
\newcommand{\dbraket}[1]{\langle\!\langle #1 \rangle\!\rangle}
\begin{document}

\title{Renormalization group improved predictions for $t\bar{t}W^\pm$ production at hadron colliders}
\author{Hai Tao Li}
\affiliation{School of Physics and State Key Laboratory of Nuclear Physics and Technology, Peking University, Beijing 100871, China}
\author{Chong Sheng Li}
\email{csli@pku.edu.cn}
\affiliation{School of Physics and State Key Laboratory of Nuclear Physics and Technology, Peking University, Beijing 100871, China}
\affiliation{Center for High Energy Physics, Peking University, Beijing 100871, China}
\author{Shi Ang Li}
\affiliation{School of Physics and State Key Laboratory of Nuclear Physics and Technology, Peking University, Beijing 100871, China}

\begin{abstract}

We study the factorization and resummation of the $t\bar{t}W^\pm$ production at hadron colliders. The cross section in the threshold limit can be factorized into a convolution of hard and soft functions and parton distribution functions with the soft-collinear effective theory. We calculate the next-to-leading order soft function for the associated production of the heavy quark pair  and colorless particle, and we perform the resummation calculation with the  next-to-next-to-leading logarithms accuracy. Our results show that the resummation effects reduce the dependence of the cross section on the scales significantly and  increase the total cross section by $7-13\%$ compared with NLO QCD results.

\end{abstract}
\maketitle


\section{Introduction}
\label{sec:introduc}

The top quark is the most massive known particle. And due to its large mass, the top quark plays a special role in the Standard Model (SM). The top quark pair  production in association with the vector boson is one of the key processes to measure top quark properties.
Furthermore, these processes can also lead to final states that contain same-sign leptons, which are rare events in the SM, and an important background in searches for new physics, such as supersymmetry and the Randall-Sundrum model. Recently the charge asymmetry between the $t$ and $\bar t$ quarks has been studied with the QCD next-to-leading order (NLO)  accuracy~\cite{Maltoni:2014zpa}. The authors pointed out that the $t\bar{t}W^\pm$ events provide larger charge asymmetry with respect to $t\bar{t}$ and $t\bar{t}Z$ production, which is also sensitive to the new physics.
On the other hand, the LHC already has the ability to distinguish the $t\bar{t}V(W^\pm,~Z)$ signals from the SM backgrounds~\cite{Chatrchyan:2013qca,Khachatryan:2014ewa}. With increasing  precision of  measurements in the near future, it is necessary to make more accurate predictions for these processes.

The calculations of the $t\bar{t}V$ production are  similar to the massive bottom quark pair production associated with a vector boson. The QCD NLO  corrections to the massive $b\bar{b}$ and $W$ or $Z$ boson associated production have been investigated in the last several years~\cite{Cordero:2006sj,Cordero:2007ce,FebresCordero:2008ci,Badger:2010mg}.
The NLO corrections to the $t\bar{t}W^\pm$ and $t\bar{t}Z$ production were also studied in Refs.~\cite{Lazopoulos:2008de,Lazopoulos:2007bv,Hirschi:2011pa,Kardos:2011na}. And the NLO corrections to the production and decay of $t\bar{t}W^\pm$ are available \cite{Badger:2010mg}. Recently,  the computations for both the $t\bar{t}W^\pm$ and $t\bar{t}Z$ production at the QCD NLO level with parton shower were presented in Ref.~\cite{Garzelli:2012bn}.
 However, the cross sections for the $t\bar{t}W^\pm$ production still suffer from large uncertainties, about $\pm10\%$, caused by the renormalization and factorization scales, even including QCD NLO effects.
As we know, the perturbative QCD calculations lead to the functions singular at the edge of the phase space and produce logarithms. In general, these logarithms can induce large corrections and bring large uncertainties in the fixed-order calculations. To improve the theoretical predictions, these logarithms must be resummed to all orders.

In this paper, we study the threshold resummation for the $t\bar{t}W^\pm$  production at hadron colliders, within the framework of the soft-collinear effective theory (SCET)~\cite{Bauer:2000ew,Bauer:2000yr,Bauer:2001ct,Bauer:2001yt}.
In general, there are several threshold parameters used in soft gluon resummation. These parameters vanish in the limit where real gluon emission is soft or collinear. As for top quark pair production, the threshold parameter could be $\beta_t=\sqrt{1-\frac{4 m_t^2}{\hat{s}}}$, $(1-z)=1-M^2/\hat{s}$ and $s_4=\hat s + \hat t+\hat u$, which correspond to the production threshold kinematics~\cite{Moch:2008qy}, pair-invariant mass (PIM) kinematics~\cite{Ahrens:2010zv,Cacciari:2011hy} and single particle inclusive(1PI) kinematics~\cite{Kidonakis:2010dk,Ahrens:2011mw}, respectively. The production threshold kinematics cannot be directly used in the resummation for $t\bar{t}W^\pm$ production and we do not consider this case in this paper. As shown in Ref.~\cite{Ahrens:2011mw}, the resummation predictions in 1PI kinematics are susceptible to large power corrections at the LHC, and the results in PIM kinematics seem more reliable for top quark pair production.
Therefore, for $t\bar{t}W^\pm$ production at the LHC,  we choose the PIM kinematics threshold where $(1-z)=1-M^2/\hat{s}\to 0$ with $M$ the invariant mass of the $t\bar{t}W^\pm$. In this case, the logarithms  $\alpha_s^n[\ln^m(1-z)/(1-z)]_+$ with $m\le 2n-1$ are induced in the perturbative expansion of the strong coupling constant  $\als$, which could spoil the convergence of the expansion. These logarithms can be resummed to to all orders in $\als$ using renormalization group method.
%
In general, in this threshold limit, the cross section can be factorized into a convolution of hard and soft functions as~\cite{Idilbi:2005ky,Becher:2007ty,Ahrens:2008qu,Ahrens:2008nc,Mantry:2009qz,
Zhu:2009sg,Idilbi:2009cc,Yang:2006gs,Zhu:2010mr,Shao:2013bz,Zhan:2013sza,Wang:2014mqt}. The cross section for $t\bar{t}W^\pm$ production can be written as
\begin{align}
    \sigma =  \mathrm{Tr}(\bm{H} \otimes \bm{S})\otimes  f_{N_1} \otimes f_{N_2}  \, ,
\end{align}
where $\bm{H}$, $\bm{S}$, and $f_{N}$ are the hard function, soft function and parton distribution function, respectively. The short distance information is encoded into the Wilson coefficients, which is described by the hard function. The soft function contains all the effects coming from emitting soft gluons by the colored initial and final states. The hard function and soft functions can be calculated order by order in QCD at the hard scale and soft scale, respectively. Then all the scales evolve to the common scale to resum the large logarithms. In this paper, we calculate the soft function at the NLO level, and then perform the threshold resummation with the NLO+NNLL accuracy.

This paper is structured as follows. In the following section, we briefly derive the factorization formula in the threshold region for $t\bar{t}W^\pm$ production. In Sec.~\ref{sec:hard&soft} we present the results of the NLO hard and soft functions.  Then we show the renormalization group equations for the hard and soft functions in Sec.~\ref{sec:RG evolution}. By solving these RG equations, the final resummation formula is given in this section.  In Sec.~\ref{sec:numer}, we discuss the numerical results for the resummation at the LHC. Finally, we conclude in Sec.~\ref{sec:conclude}.

\section{Factorization formula}
\label{sec:factoriz}

In this section, we briefly show the factorization formula for $t\bar{t} W^\pm$ production based on SCET and heavy-quark effective theory (HQET)~\cite{Isgur:1989vq}. SCET is developed as a useful tool to deal with soft and collinear radiations. And here HQET is used to describe the interactions between the soft gluon and the top quark pair in the final states.
Because the $W$ boson is colorless, the factorization procedure shares some  similarities with the case of the top pair production, which has been discussed in detail in Refs.~\cite{Ahrens:2010zv,Li:2013mia}.

We consider the process
\begin{align}
   N_1(P_1)+N_2(P_1)\to t (p_3) + \bar{t} (p_4) + W^{\pm} (p_5) + X(P_X),
\end{align}
where $N_1$ and $N_2$ are the incoming hadrons and $X$ is an inclusive hadronic final state. At the LO, the process is
\begin{align}
      q(p_1) + \bar{q}(p_2) \to t(p_3)  + \bar{t} (p_4) + W^\pm(p_5)\, ,
\end{align}
where $p_1=x_1 P_1$ and $p_2=x_2 P_2$. It is convenient to introduce the following kinematic invariants:
\begin{gather}
    s=(P_1+P_2)^2\ , \qquad \hat{s}=(p_1+p_2)^2\ ,  M^2=(p_3+p_4+p_5)^2\, ,
     \nn \\
    \qquad s_{ij} = (p_i+p_j)^2\, ,  \qquad  \tilde{s}_{ij}=2 p_i \cdot p_j\, , \qquad
    z=\frac{M^2}{\hat{s}}, \qquad \tau = \frac{M^2}{s}\,  ,
\end{gather}
The threshold limit we are interested in is $(1-z)\to 0$, where
\begin{align}
     \hat{s},\ \tilde{s}_{ij},\ M^2 \gg \hat{s} (1-z)^2 \gg \Lambda_{QCD}^2.
\end{align}
In this region only the soft radiations in the final state are allowed and $(1-z)^2 \hat{s}$ defines the soft scale. For later convenience, we introduce two lightlike vectors $n$ and $\bar{n}$ along the directions of the colliding partons. With these two vectors, any four-vector can be written  as
\begin{align}
        k^\mu=n\cdot k \frac{\bar{n}^\mu}{2}+\bar{n}\cdot k \frac{n^\mu}{2}+k_\perp^{\mu}
        = k^+ \frac{\bar{n}^\mu}{2}+k^- \frac{n^\mu}{2}+k_\perp^{\mu}=(k^+,k^-,\vec{k}_\perp)\, .
\end{align}
Therefore, the momenta of the initial parton can be written as $p_1^\mu=p^- n^\mu/2$ and $p_2=p^+ n^\mu/2$. The momenta of the top quark pair in the final states are $p_i^\mu=m_t v_i^\mu+k_i^\mu$ ($i$=3,4) where $v_i^2=1$ and $k_i^\mu$ is the off-shell momentum due to the soft gluon emissions.

It is similar to the factorization procedure of threshold resummation for top quark pair production in the quark-antiquark annihilation channel~\cite{Ahrens:2010zv,Li:2013mia}. In order to derive the explicit expression of the factorization formula in SCET,  we start from the effective Hamiltonian for $t\bar{t}W$ production, which can be written as
\begin{align}
  \label{eq:hamiltonian}
  \mathcal{H}_{\text{eff}}(x) = \sum_{I,m} \int dt_1 dt_2 \, e^{im_t(v_3+v_4) \cdot x} \tilde{C}_{Im}(t_1,t_2) \mathcal{W}_\mu(x) O_{Im}^{\mu}(x,t_1,t_2)\ ,
\end{align}
where the index $I$ and $m$ label different color structures and Dirac structures, respectively. $\mathcal{W}_\mu(x)$  is the field operator of the $W$ boson and $\tilde{C}(t_1,t_2)$ are Wilson coefficients arising from matching the renormalized Green's functions in QCD  with the operator $O_{Im}^{\mu}(x,t_1,t_2)$ in SCET.  In the momentum space, this Wilson coefficients can be obtained by performing integral over $t_1$ and $t_2$
\begin{align}
   C_{Im}(\mu)=\int dt_1 dt_2 e^{-i t_1 \bar{n}\cdot p_1 - i t_2 n \cdot p_2} \tilde{C}_{Im}(t_1,t_2),
\end{align}
Here we suppress the arguments for the Wilson lines because of many independent kinematic variables involved.

The effective operators in SCET in Eq.~(\ref{eq:hamiltonian}) are given by
\begin{align}
  O_{Im}^{\mu}(x,t_1,t_2) = \sum_{\{a\},\{b\}}  (c_I)_{\{a\}} \, [O_m^h(x)]^{b_3b_4} \, [O_m^{c, \mu}(x,t_1,t_2)]^{b_1b_2} \, [O^s(x)]^{\{a\},\{b\}} \, ,
\end{align}
with
\begin{gather}
[O^{c,\mu}_m(x,t_1,t_2)]^{b_1b_2} = \bar{\chi}_{\bar{n}}^{b_2}(x+t_2n) \, \Gamma^{\mu}_m \, \chi_n^{b_1}(x+t_1\bar{n}) \, , \quad [O_m^h(x)]^{b_3b_4} = \bar{h}_{v_3}^{b_3}(x) \, \Gamma'_m \, h_{v_4}^{b_4}(x)  \nn
  \\
  [O^s(x)]^{\{a\},\{b\}} = [S^\dagger_{v_3}(x)]^{b_3a_3} \, [S_{v_4}(x)]^{a_4b_4} \, [S^{\dagger}_{\bar{n}}(x)]^{b_2a_2} \, [S_n(x)]^{a_1b_1} \, .
  \label{eq:qqoperators}
\end{gather}
In the above,  $\chi_n$, $h_v$  are gauge-invariant fields for collinear quarks,  heavy quarks in SCET and HQET, respectively. The superscripts $\mu$ is the Lorentz index. And the indices $a_i$ and $b_i$ with $i=1,2,3,4$ are color indices for the initial and final quarks. And $\Gamma^\mu_m$, $\Gamma'_m$ are combinations of Dirac matrices and the external momentum $n$, $\bar{n}$, $v_3$ and $v_4$.
In the above equation, we have used the soft Wilson lines which are defined as~\cite{Chay:2004zn,Korchemsky:1991zp}
\begin{align}
  [S_n(x)]^{ab} &= \mathcal{P} \exp \left( ig\int_{-\infty}^0 dt \, n \cdot A_s^c(x+tn) \, t^c_{ab} \right)\, ,
  \nn \\
   [S_{v_3}(x)]^{ab} &= \mathcal{P} \exp \left( -ig\int_{0}^{\infty} dt \, v_3 \cdot A_s^c(x+tv_3) \, t^c_{ab} \right) \, .
\end{align}
$[S_{\bar{n}}(x)]^{ab}$  and $[S_{v_4}(x)]^{ab}$  are similar to $[S_n(x)]^{ab}$  and $[S_{v_3}(x)]^{ab}$, respectively, and we do not show them. To suppress the color indices, it is convenient to introduce the color-space formalism~\cite{Catani:1996jh,Catani:1996vz}.
The basis in the color space is denoted by  $c_I$, which we choose as
\begin{align}
  \left(c_1\right)_{\{a\}} = \delta_{a_1a_2} \, \delta_{a_3a_4} \, , \quad \left(c_2\right)_{\{a\}} = t_{a_1a_2}^c \, t_{a_3a_4}^c \, .
  \label{eq:colorbasis}
\end{align}
Using these bases, we define the vectors of Wilson coefficients as
\begin{align}
  \Ket{C_m} = \sum_{I} C_{Im} \ket{c_I} \, .
\end{align}

Due to the fact that the fields in different sectors of the effective theory do not interact with each other, after absorbing the corresponding interactions into Wilson lines, the partonic differential cross section can be expressed as
\begin{align}
\label{eq:dshat}
  d\hat\sigma &= \frac{1}{2 \hat{s}}  \frac{d^3 \vec{p}_3}{ (2\pi)^3 2 E_3}\frac{d^4 \vec{p}_4}{ (2\pi)^3 2 E_4}
  \frac{d^3 \vec{p}_5}{ (2\pi)^3 2 E_5} (2\pi)^4\delta^{(4)}(p_1+p_2-p_3-p_4-p_5-p_s)
  \nn \\ & \times \frac{1}{4 N_C^2} \sum_{m,m^\prime} |C_m \rangle \langle C_{m^\prime}|
   \dbraket{O_m}^\dagger \dbraket{O_{m^\prime}} \langle 0| \bar{\bm{T}}[\bm{O}^{s\dagger}(x)]
   \bm{T}[\bm{O}^s(0)]| 0 \rangle\ ,
\end{align}
where $p_s$ denotes the four-momentum of soft radiations. Here, we introduce the symbol $\dbraket{O_m}$ which is defined as
\begin{align}
 \dbraket{O_m} = \langle t(p_3)\bar{t}(p_4)|\mathcal{O}^h_m(0,0,0) \epsilon_{\mu}(p_5) \mathcal{O}^{c,\mu}_m(0,0,0) | u(p_1)\bar{d}(p_2)\rangle \, .
\end{align}
The hard function is a matrix in the color space, which is  given by
\begin{align}
   \bm{H}(\mu) &= \frac{1}{4 N_C} \sum_{m,m^\prime} |C_m \rangle \langle C_{m^\prime}|
   \dbraket{O_m}^\dagger \dbraket{O_{m^\prime}}\ .
\end{align}
Then we define the soft function in the position space as the vacuum expectation value of Wilson loops,
\begin{align}
   \nn \\
   \bm{W}(x,\mu)&=\frac{1}{N_C} \langle 0| \bar{\bm{T}}[\bm{O}^{s\dagger}(x)]
   \bm{T}[\bm{O}^s(0)]| 0 \rangle\,  ,
\end{align}
which is also a matrix in color space.
And in the momentum space the soft function is given by
\begin{align}
  \label{eq:softmom}
  \bm{S}(\sqrt{\hat{s}}(1-z),\mu) = \sqrt{\hat{s}} \int \frac{dx_0}{4\pi}
  \, e^{i\sqrt{\hat{s}}(1-z)x_0/2} \, \bm{W}(x_0,\vec{x}=0,\mu) \, .
\end{align}

Using the following identity
\begin{align}
    1 = \int d^4q \, dM^2 \, \delta^{(4)}(q-p_3-p_4-p_5) \, \delta(M^2-q^2) \, ,
\end{align}
 Eq.~(\ref{eq:dshat}) can be  expressed as
\begin{align}
    d\hat\sigma =&
    \frac{1}{2 \pi}  \frac{1}{2 \hat{s}} d\Pi_3 dM \int dx^0 e^{i x^0 (p_1^0+p_2^0-q^0)}
    \mathrm{Tr} [\bm{H}(\mu)\bm{W}(x,\mu_f)]  \delta(M^2-q^2)
    \nn \\
    =& \frac{1}{\hat{s} M} d\Pi_3 dM \mathrm{Tr} [\bm{H}(\mu)\bm{S}(\sqrt{\hat{s}}(1-z),\mu)] \delta(M^2-q^2)
\end{align}
where $d\Pi_3$ are
\begin{align}
      d\Pi_3 = \frac{d^3 \vec{p}_3}{ (2\pi)^3 2 E_3}\frac{d^4 \vec{p}_4}{ (2\pi)^3 2 E_4}
  \frac{d^3 \vec{p}_5}{ (2\pi)^3 2 E_5} (2\pi)^4\delta^{(4)}(q-p_3-p_4-p_5)\, .
\end{align}
Then the differential cross section at hadron colliders is
\begin{align}
    \frac{d\sigma}{dM d\Pi_3} = \frac{1}{s} \sum_{i=q,\bar{q}}\int_{\tau}^1 \frac{dz}{z} \ff_{i\bar{i}}(\tau/z,\mu) C(z,\mu)\ ,
\end{align}
where  $C(z,\mu)$ is the hard-scattering kernel, which is given by
\begin{align}
  C(z,\mu) = \mathrm{Tr} \big[ \bm{H}(\mu)  \bm{S}(\sqrt{\hat{s}}(1-z),\mu) \big] \, .
\end{align}
The variable $\ff_{i\bar{i}}$ is the convolution of the PDFs, which is defined as
\begin{align}
         \ff_{i\bar{i}}(y,\mu_f) = \int_y^1 \frac{dx}{x} f_{i/N_1}(x,\mu)f_{\bar{i}/N_2}(y/x,\mu)\, .
\end{align}

\section{The hard function and soft function}
\label{sec:hard&soft}

In this section we first summarize the results of the hard function and then show the NLO soft function. The hard function is the absolute value squared of the Wilson coefficients of the operators, which can be obtained by matching the full theory onto SCET.
The perturbation expansion of the hard function can be written as
\begin{align}
\label{eq:exp_hard}
  \bm{H} = \alpha_s^2 \, \frac{1}{N_C} \left( \bm{H}^{(0)} + \frac{\alpha_s}{4\pi}
    \bm{H}^{(1)} + \ldots \right) .
\end{align}
The LO hard function $\bm{H}^{(0)}$ is simple to calculate. For the NLO hard function $\bm{H}^{(1)}$, the external particles are chosen on shell, and the loop integrals in SCET vanish in dimensional regularization  because  scaleless. The NLO hard function is exactly the same with the renormalized virtual corrections for the $t\bar{t}W^{\pm}$ production. The NLO calculations for the $t\bar{t}W^\pm$ production have also been performed in  Refs.~\cite{Badger:2010mg, Hirschi:2011pa}. And the NLO hard function we need can be extracted from  the MadLoop~\cite{Hirschi:2011pa}, which makes use of CutTools~\cite{Ossola:2007ax} and OneLoop~\cite{vanHameren:2009dr}.
As a cross check, we use the anomalous dimension of the hard function~\cite{Becher:2009kw,Ferroglia:2009ep,Ferroglia:2009ii} to predict the divergence and the scale dependent terms in the NLO hard function. And we find that these terms are consistent with those from MadLoop.

Now we turn to the soft function. It is convenient to work with the Laplace transformed function of the soft function.  With the definition in Eq.~(\ref{eq:softmom}), the soft functions can be given by
\begin{align}
  \label{eq:stilde}
  \tilde{\bm{s}}(L, \mu) &= \frac{1}{\sqrt{\hat s}} \int_0^\infty d\omega
  \, \exp \left( -\frac{\omega}{e^{\gamma_E}\mu e^{L/2}} \right)
  \bm{S}(\omega, \mu) \nonumber
  \\[-2mm]
  &= \bm{W} \bigg( x_0 = \frac{-2i}{e^{\gamma_E}\mu e^{L/2}}, \mu \bigg) \, .
\end{align}
The perturbation expansion of the soft function can be written as
\begin{align}
     \tilde{\bm{s}}(L,\mu) = \tilde{\bm{s}}^{(0)}(L,\mu)+\frac{\als}{4\pi}\tilde{\bm{s}}^{(0)}(L,\mu)\, .
\end{align}
At the LO, the soft function is independent of the $L$ and $\mu$ , which is
\begin{align}
\tilde{\bm{s}}^{(0)}_{IJ} =& \Braket{c_I|c_J}/N_C\, \nn
\end{align}
or
\begin{align}
  \tilde{\bm{s}}^{(0)} =&
  \begin{pmatrix}
    N_C & 0
    \\
    0 & \frac{C_F}{2}
  \end{pmatrix} \, .
\end{align}
\begin{figure}[t!]
  \includegraphics[width=0.8\textwidth]{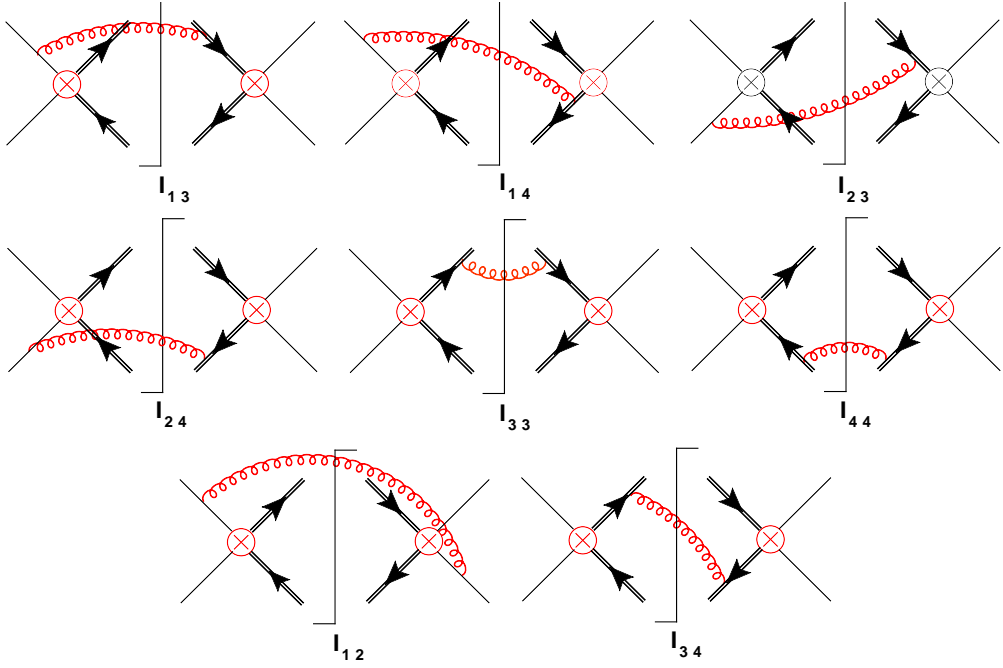}
  \vspace{-2ex}
  \caption{\label{fig:soft_funcs} The diagrams contributing to the soft functions. The single line and double line represent the Wilson line for initial-state quarks and final state top quark pair and the red one for the soft gluon.}
\end{figure}

Figure~\ref{fig:soft_funcs} shows the diagrams for the NLO soft function, which can be calculated using the operator definition in SCET or using the eikonal approximation in the full theory. The bare soft function can be written as
\begin{align}
  \bm{W}^{(1)}_{\text{bare}}(\epsilon,x_0,\mu) = \sum_{i,j} \, \bm{w}_{ij} \,
  \mathcal{I}_{ij}(\epsilon,x_0,\mu) \, ,
\end{align}
where the $\bm{w}_{ij}$ represents the  NLO color matrices, which is defined as
\begin{align}
     (\bm{w}_{ij})_{IJ} = \frac{1}{N_C} \Braket{c_I|\bm{T}_i \bm{T}_j|c_J}\, .
\end{align}
These color matrices are
\begin{align}
  \label{eq:qTT}
  \bm{w}_{12} = \bm{w}_{34} &= -\frac{C_F}{4N_C}
  \begin{pmatrix}
    4N_C^2 & 0
    \\
    0 & -1
  \end{pmatrix}
  , \nonumber
  \\
  \bm{w}_{33} = \bm{w}_{44} &= \frac{C_F}{2}
  \begin{pmatrix}
    2N_C & 0
    \\
    0 & C_F
  \end{pmatrix}
  , \nonumber
  \\
 \bm{w}_{13} = \bm{w}_{24}
 &=- \frac{C_F}{2}
  \begin{pmatrix}
    0 & 1
    \\
    1 & 2C_F - \frac{N_C}{2}
  \end{pmatrix}
  , \nonumber
  \\
  \bm{w}_{14} = \bm{w}_{23} &= \frac{C_F}{2N_C}
  \begin{pmatrix}
    0 & N_C
    \\
    N_C & 1
  \end{pmatrix}
  .
\end{align}
The integrals $\mathcal{I}_{ij}$ are defined as
\begin{align}
  \label{eq:softintegrals}
  \mathcal{I}_{ij}(\epsilon,x_0,\mu) = -\frac{(4\pi\mu^2)^\epsilon}{\pi^{2-\epsilon}} \,
  v_i \cdot v_j \int d^dk \, \frac{e^{-ik^0x_0}}{v_i \cdot k \, v_j \cdot k} \, (2\pi) \,
  \delta(k^2) \, \theta(k^0) \, ,
\end{align}
where $v_i$ is the velocity of the corresponding particle $i$. It is obvious that the integrals $\mathcal{I}_{ij}$ are symmetric in the indices $i$ and $j$.  The integrals $\mathcal{I}_{11}$, $\mathcal{I}_{22}$ vanish because of massless external particles. And the integral $\mathcal{I}_{12}$ and $\mathcal{I}_{33}$ are the same with those in Ref.~{\cite{Ahrens:2010zv}}. The other integrals need to be recalculated here due to the difference between the  kinematic conditions for  $t\bar{t}$ and $t\bar{t}W^\pm$ production.  The non vanishing integrals are collected as follows
\begin{align}
  \mathcal{I}_{12} =& -\left( \frac{2}{\epsilon^2} + \frac{2}{\epsilon} L_0 + L_0^2 +
    \frac{\pi^2}{6} \right)\, ,
    \nonumber  \\
  \mathcal{I}_{33} =& \frac{2}{\epsilon} + 2L_0 - \frac{2}{\beta_3} \ln \frac{1-\beta_3}{1+\beta_3} \, , \nonumber
  \\
  \mathcal{I}_{34} =& -\frac{1+x_{34}^2}{1-x_{34}^2} \left[ \left( \frac{2}{\epsilon} + 2L_0
    \right) \ln x_{34} - \ln^2x_{34} + 4\ln x_{34} \ln(1-x_{34}) + 4\mathrm{Li}_2(x_{34}) - \frac{2\pi^2}{3}
  \right] , \nonumber
  \\
  \mathcal{I}_{13} =&-\frac{1}{\epsilon^2}
   -\frac{1}{\epsilon} \left[ L_0- \ln\frac{\tilde{s}_{13}^2}{s m_t^2} \right]
   -\left[\frac{1}{2} \left( L_0- \ln\frac{\tilde{s}_{13}^2}{s m_t^2} \right)^2
     \nn \right.  \\  & \left.
   +2\mathrm{Li}_2\left(\frac{\beta_3 (1+\cos\theta_3)}{1+\beta_3}\right) +
    2\mathrm{Li}_2\left(-\frac{\beta_3 (1-\cos\theta_3)}{1-\beta_3}\right) + \frac{\pi^2}{12} \right]\ ,
    \nn   \\
   \mathcal{I}_{23} =& \mathcal{I}_{13}\ (\cos\theta_3 \to -\cos\theta_3) \, ,\qquad
   \mathcal{I}_{14} = \mathcal{I}_{13}\ (\beta_{3} \to \beta_{4}, \theta_3 \to \theta_4) \, ,
   \nn \\
   \mathcal{I}_{24} =& \mathcal{I}_{14}\ (\cos\theta_4 \to -\cos\theta_4)\ ,\qquad
   \mathcal{I}_{44} = \mathcal{I}_{33}\ (\beta_3 \to \beta_4)\, .
   \end{align}
where $\theta_3$ and $\theta_4$ are the angle between the top quark and anti-top quark  momentum and z axes , respectively, and
\begin{align}
  L_0 = \ln \bigg( -\frac{\mu^2x_0^2e^{2\gamma_E}}{4} \bigg) \, , \qquad
  \beta_{i}= \frac{|\vec{p_i}|}{p_i^0} \, , \qquad
  \beta_{t\bar{t}} = \sqrt{1-\frac{4m_t^2}{s_{34}}}\, \qquad \,
  x_{34} = \frac{1-\beta_{t\bar{t}}} {1+\beta_{t\bar{t}}} \, .
\end{align}

If we remove the momentum of the $W$ boson, .i.e. set $p_1+p_2=p_3+p_4$, the soft function here turns into that for  top pair production, which is consistent with the results from Ref.~\cite{Ahrens:2010zv}. Furthermore, the real emissions of NLO corrections in the soft limit were present in Ref.~\cite{Cordero:2006sj,Cordero:2007ce}, where the NLO corrections to the massive bottom quark pair production in association with $W^\pm$ boson are calculated. As a check, we transform the soft function to the momentum space, and find they are consistent with the results in Ref.~\cite{Cordero:2007ce}. The soft functions presented here can be used to all the processes of heavy quark pair production associated with colorless particle at the hadron colliders.

\section{Renormalization Group evolution and the threshold resummation}
\label{sec:RG evolution}

With SCET, the resummation of large logarithms is achieved by evolving the hard function and soft function from the hard scale $\mu_h$ and soft scale $\mu_s$ to the factorization scale $\mu_f$. In this section, we will introduce the RG evolution of the hard and soft function, respectively. The resummation formula and its expansion to fixed order are also presented below.

\subsection{RG evolution and resummation}
The RG evolution of the hard functions is similar to the one for top quark pair production, which is given by
\begin{align}
  \label{eq:RG_H}
  \frac{d}{d\ln\mu} \bm{H}(\mu) &= \bm{\Gamma}_H (\mu) \, \bm{H}(\mu)
   + \bm{H}(\mu) \, \bm{\Gamma}_H^{\dagger}(\mu) \, .
\end{align}
The anomalous dimension $\bm{\Gamma}_H$ can be obtained by applying the results
derived in Ref.~\cite{Becher:2009kw,Ferroglia:2009ep,Ferroglia:2009ii},  which can be written as
\begin{align}
   \bm{\Gamma}_H =& \left[C_F \gamma_{\rm{cusp}}(\als) \left(\ln\frac{M^2}{\mu^2}-i\pi \right)
         + C_F \gamma_{\rm{cusp}}(\beta_{34},\als) + 2\gamma^q (\als) + 2\gamma^Q (\als) \right] \bm{1}
                  \nn \\ &
         + \frac{N_C}{2} \left[ \gamma_{\rm{cusp}}(\als)
         \left(  \frac{1}{2} \ln \frac{\tilde{s}_{13}^2}{\hat{s} m^2_t} + \frac{1}{2} \ln \frac{\tilde{s}_{24}^2}{\hat{s} m^2_t}
         + i\pi\right)
         - \gamma_{\rm{cusp}} (\beta_{34},\als) \right]
         \begin{pmatrix}
          0 & 0
          \\
          0 & 1
         \end{pmatrix}
                  \nn \\ &
         + \gamma_{\rm{cusp}} (\als) \left( \ln \frac{\tilde{s}_{13}}{\tilde{s}_{23}}- \ln \frac{\tilde{s}_{14}}{\tilde{s}_{24}} \right)
         \left[
         \begin{pmatrix}
          0 & \frac{C_F}{2 N_C}
          \\
          1 & -\frac{1}{N_C}
         \end{pmatrix}
         +\frac{\als}{4\pi} g(\beta_{34})
         \begin{pmatrix}
          0 & \frac{C_F}{2}
          \\
          -N_C & 0
         \end{pmatrix}
         \right]\, ,
\end{align}
where $\gamma_{\text{cusp}}$, $\gamma^q$ and $\gamma^Q$ are the cusp anomalous dimension, light quark field anomalous dimension, and heavy quark field anomalous dimension, respectively. The results for these anomalous dimensions can also be found in Refs.~\cite{Becher:2009kw,Ferroglia:2009ep,Ferroglia:2009ii}.

The solution to Eq.~(\ref{eq:RG_H}) is given by
\begin{align}
  \bm{H}(\mu) = \bm{U}_{H}(\mu) \, \bm{H}(\mu_h) \, \bm{U}_H^{\dagger}(\mu_h,\mu) \, ,
\end{align}
where $\bm{U}_{H}$ is
\begin{align}
\label{eq:rg_UH}
  \bm{U}_{H}( \mu_h,\mu) &= \exp \bigg[ 2S(\mu_h,\mu)-a_{\Gamma}(\mu_h,\mu) \left( \ln \frac{M^2}{\mu_h^2}-i\pi\right) \bigg]   \bm{u}(\mu_h,\mu) \, .
\end{align}
The functions $S(\mu_h,\mu)$ and $a_{\Gamma}$ are defined as
\begin{align}
  S(\mu_h,\mu) = - \int_{\als(\mu_h)}^{\als(\mu)} d\alpha \, \frac{\Gamma_{\text{cusp}}^{F}(\alpha)}{\beta(\alpha)} \int_{\als(\mu_h)}^\alpha \frac{d\alpha'}{\beta(\alpha')} \, , \quad a_{\Gamma}(\mu_h,\mu) = - \int_{\als(\mu_h)}^{\als(\mu)} d\alpha \, \frac{\Gamma_{\text{cusp}}^{F}(\alpha)}{\beta(\alpha)} \, ,
\end{align}
where $\beta(\alpha)$ is the QCD $\beta$ function and $\Gamma_{\text{cusp}}^F=C_F\gamma_{\text{cusp}}$. The matrix $\bm{u}$ in Eq.~(\ref{eq:rg_UH}) is
\begin{align}
  \label{eq:uh}
  \bm{u}(\mu_h,\mu) = \mathcal{P} \exp \int_{\als(\mu_h)}^{\als(\mu)} \frac{d\alpha}{\beta(\alpha)} \, \bm{\gamma}_{h}(\mu_h,\mu) \, ,
\end{align}
 The above equation can be solved through the method introduced in Refs.~\cite{Buras:1991jm,Buchalla:1995vs}. The matrix $\bm{\gamma}_{h}(\als)$ can be obtained from the following equation:
\begin{align}
  \bm{\Gamma}_{H} =  \Gamma^F_{\text{cusp}}(\als) \left( \ln\frac{M^2}{\mu^2} - i\pi \right) \bm{1} + \bm{\gamma}_{h}(\als) \, .
\end{align}
Finally, the RG improved hard function is
\begin{align}
  \bm{H}( \mu_h,\mu) &=\exp \bigg[ 4S(\mu_h,\mu)-2a_{\Gamma}(\mu_h,\mu) \left( \ln \frac{M^2}{\mu_h^2}-i\pi\right) \bigg]  \bm{u}( \mu_h,\mu) \, \bm{H}( \mu_h) \, \bm{u}_h^{\dagger}(\mu_h,\mu) \, .
\end{align}

The cross section at the threshold region is independent of the factorization scale, and then we have
\begin{align}
\label{eq:tota_scale_indep}
    \frac{d}{d\ln \mu} \mathtt{Tr}[\bm{H}(\mu) \bm{S}(\omega, \mu)] \otimes \ff =0\ .
\end{align}
where $\ff$ is the convolution of the parton distribution functions. The RG equation of $\ff$ is
\begin{align}
    \frac{d}{d\ln \mu} \ff(y,\mu) = 2 \int_y^1 P(x)\ff(y/x,\mu)\ ,
\end{align}
where
\begin{align}
      P(x) = \frac{2 \Gamma^F_{\text{cusp}}(\als)} {(1-x)_+} + 2\gamma^\phi (\als)\delta(1-x)\ .
\end{align}
From Eq.~(\ref{eq:tota_scale_indep}), we find that the momentum-space soft function obeys the following integro-differential equation:
\begin{align}
  \frac{d}{d\ln\mu} \bm{S}(\omega, \mu) &= - \left[
    2\Gamma_{\text{cusp}}^F(\alpha_s) \ln\frac{\omega}{\mu} +
    \bm{\gamma}_s^{\dagger}(\alpha_s) \right]
  \bm{S}(\omega,\mu)
    \nn \\ & \hspace{-3em}
  - \bm{S}(\omega,\mu) \left[
    2\Gamma^{F}_{\text{cusp}}(\alpha_s) \ln\frac{\omega}{\mu} +
    \bm{\gamma}_{s}(\alpha_s) \right]
 - 4\Gamma_{\text{cusp}}^{F}(\alpha_s) \int_0^\omega d\omega' \,
  \frac{\bm{S}(\omega',\mu)-\bm{S}(\omega, \mu)}
  {\omega-\omega'} \, ,
\end{align}
where $ \bm{\gamma}_s(\alpha_s) = \bm{\gamma}_h(\alpha_s) +  2\gamma^{\phi}(\alpha_s) \, \bm{1} \,$. With the Laplace transformation, the nonlocal evolution equation for the soft function is turned into a local one. The evolution equation after the Laplace transformation is given by
\begin{align}
  \label{eq:Sev}
  \frac{d}{d\ln\mu} \, \tilde{\bm{s}} \left(\ln\frac{M^2}{\mu^2},\mu\right) =
  & - \left[ \Gamma_{\text{cusp}}^{F}(\alpha_s) \ln\frac{M^2}{\mu^2} +
    \bm{\gamma}_s^{\dagger}(\alpha_s) \right] \tilde{\bm{s}}
  \left(\ln\frac{M^2}{\mu^2},\mu\right) \nonumber
  \\
  &- \tilde{\bm{s}} \left(\ln\frac{M^2}{\mu^2},\mu\right) \left[
    \Gamma_{\text{cusp}}^{F}(\alpha_s) \ln\frac{M^2}{\mu^2} +
    \bm{\gamma}_s(\alpha_s) \right] .
\end{align}
This differential equation can be solved with the same method as for the hard function. The solution in the momentum space is
\begin{align}
\label{eq:RG_soft}
  \bm{S}(\omega, \mu) =& \sqrt{\hat s}\, \exp \left[ -4S(\mu_s,\mu) +
    4a_{\gamma^\phi}(\mu_s,\mu) \right] \bm{u}^\dagger(\mu,\mu_s)
    \nn \\ & \times
  \tilde{\bm{s}}(\partial_\eta,\mu_s)
  \bm{u}(\mu,\mu_s) \, \frac{1}{\omega}
  \left(\frac{\omega}{\mu_s}\right)^{2\eta} \frac{e^{-2\gamma_E\eta}}{\Gamma(2\eta)} \, ,
\end{align}
with
\begin{align}
\eta=2a_\Gamma(\mu_s,\mu) \, .
\end{align}
Here $\mu_s$ is the soft scale where the perturbative expansion of the soft function is well behaved. This result is well defined for $\eta > 0$. When  $-1 < \eta <0$, the solution can be obtained by analytic continuation,
\begin{align}
      \int_\tau^1 dz \frac{f(z)}{(1-z)^{1-2\eta}} = \int_\tau^1 dz \frac{f(z)-f(1)}{(1-z)^{1-2\eta}}+\frac{f(1)}{2\eta} (1-\tau)^{2\eta}\ .
\end{align}

Combining the RG solutions of hard and soft function, the RG improved hard-scattering kernel is collected as follows
\begin{align}
  \label{eq:hard_kernal}
  C(z,\mu_f) =& \exp \big[ 4a_{\gamma^{\phi}}(\mu_s,\mu_f) \big]
     \nn \\  &\times \text{Tr} \bigg[ \bm{U}(\mu_h,\mu_s) \,
      \bm{H}(\mu_h) \, \bm{U}^\dagger(\mu_h,\mu_s) \tilde{\bm{s}}
     \left(\ln\frac{M^2}{\mu_s^2}+\partial_\eta,\mu_s\right) \bigg]
    \frac{e^{-2\gamma_E \eta}}{\Gamma(2\eta)} \frac{z^{-\eta}}{(1-z)^{1-2\eta}} \, .
 \end{align}

Based on the above expressions, the logarithms in the threshold region are resummed to all orders. In the following calculations,  the counting scheme  is the same with Ref.~\cite{Becher:2007ty}. To include the subleading  terms in $(1-z)$, we need to match the resummed prediction to the fixed order results. Here the prediction with NLO+NNLL accuracy is defined as
\begin{align}
\label{eq:match}
      d\sigma^{NLO+NNLL}(\mu_h,\mu_s,\mu_f) =&   d\sigma^{NNLL}|_{\mu_h,\mu_s,\mu_f}+ d\sigma^{\rm{subleading}}|_{\mu_f}
       \nn \\
       =& d\sigma^{NNLL}|_{\mu_h,\mu_s,\mu_f} + (d\sigma^{NLO}|_{\mu_f} -d\sigma^{NNLL}|_{\mu_h=\mu_s=\mu_f})\, .
\end{align}

\subsection{The NLO leading singular terms}

To explore the necessary of the resummation, we show the leading singular terms from expanding the resummed formula here and compare with the NLO results. The leading singular terms can be obtained by setting $\mu_h=\mu_s=\mu_f=\mu$ in Eq.~(\ref{eq:hard_kernal}). Thus, we take the derivatives with respect to $\eta$ and then set the limit $\eta\to 0$, which can be achieved by the replacements:
\begin{align}
     1 \to \delta(1-z), & \qquad    \qquad   \bigg(\ln\frac{M^2}{\mu_s^2}+\partial_\eta\bigg)\to 2 P_0(z) + \delta(1-z)\ln \frac{M^2}{\mu^2}\,,\nn \\
    & \hspace{-2em} \bigg(\ln\frac{M^2}{\mu_s^2}+\partial_\eta\bigg)^2  \to 4 P_1(z)+\delta(1-z)\ln^2 \frac{M^2}{\mu^2}\,,
\end{align}
where $ P_n(z)$ is plus distribution of the form
\begin{align}
     P_n(z) = \left[ \frac{1}{1-z} \ln^n \left( \frac{M^2 (1-z)^2}{\mu^2 z } \right) \right]_+\ .
\end{align}
The integration of plus distribution is defined as
\begin{align}
      \int_\tau^1 \left[ \frac{1}{1-z} \ln^n \left( \frac{M^2 (1-z)^2}{\mu^2 z } \right) \right]_+ g(z)
           =&
           \nn\\& \hspace{-15em}
           \int_\tau^1 dz \frac{g(z)-g(1)}{1-z} \ln^n \left( \frac{M^2 (1-z)^2}{\mu^2 z } \right)
           -g(1) \int_0^\tau dz \frac{1}{1-z} \ln^n \left( \frac{M^2 (1-z)^2}{\mu^2 z } \right)\ .
\end{align}

\begin{figure}[t!]
  \includegraphics[width=0.49\textwidth]{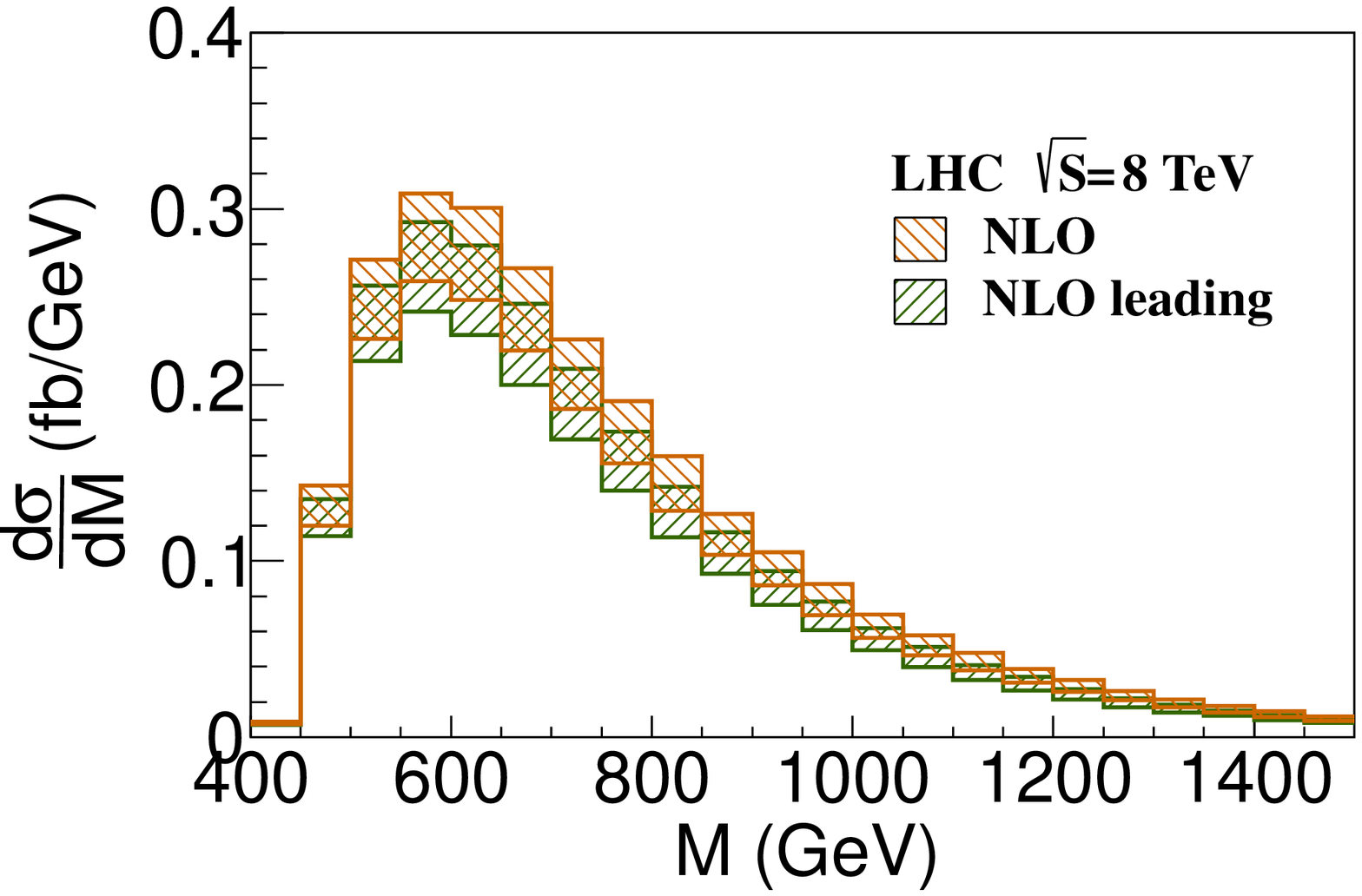}
  \includegraphics[width=0.49\textwidth]{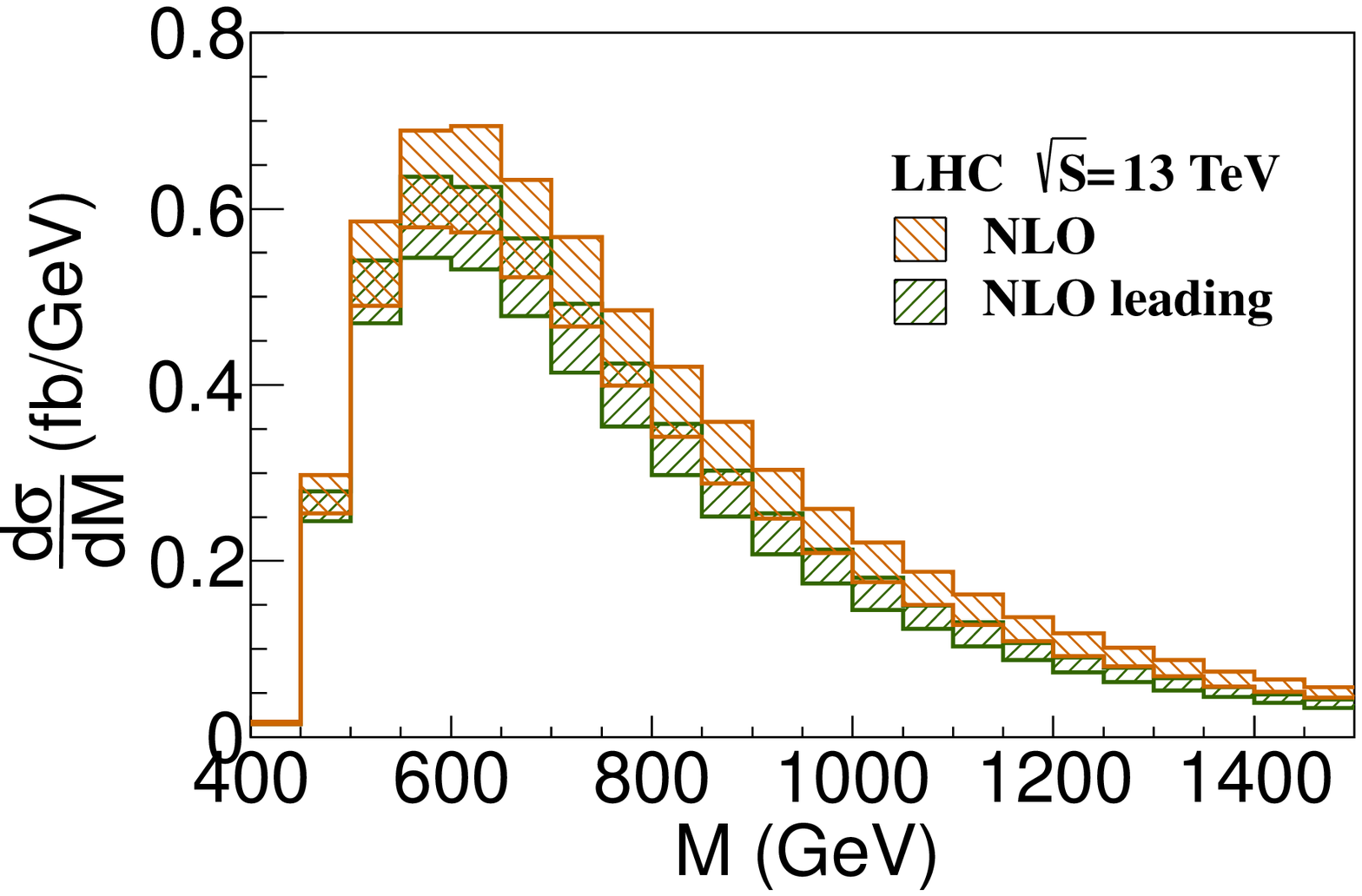}
 \vspace{-2ex}
  \caption{\label{fig:nloleading} The invariant mass distribution for the NLO leading terms and the NLO corrections for $t\bar{t}W^+$ production at the 8 and 13 TeV LHC.}
\end{figure}

Figure~\ref{fig:nloleading} shows the invariant mass distribution from the leading singular terms and the NLO results with the CT10NLO PDF \cite{Lai:2010vv}. It can be found that the leading singular terms are dominate in all the invariant mass regions and the threshold resummation effects are also important in the region $\tau \ll 1$. It is because that  the convolution of the parton distribution functions falls off fast for $\tau/z \to 1$ so that the  contributions coming from the case of  $\tau/z \ll 1$ are dominant, which is so-called dynamical threshold enhancement\cite{Becher:2007ty}. This leads that the leading singular terms contribute about $91\%$ of the NLO total cross section at the 8 TeV LHC and about $87\%$ at the 13 TeV LHC. The difference between the NLO leading singular terms and the NLO predictions come from the contribution of the subleading terms.

\section{Numerical results}
\label{sec:numer}
In this section,  we  discuss the numerical results of the resummation predictions at the LHC. The parameters are set as  $m_t=173.2$ GeV, $m_W=80.398$ GeV, and the Fermi-constant $G_F=1.166390 \times 10^{-5}$ GeV$^{-2}$. We use the CTEQ6l~\cite{Pumplin:2002vw}, CT10NLO~\cite{Lai:2010vv} and CT10NNLO~\cite{Gao:2013xoa} PDF sets as the LO, NLO and NNLO PDFs, respectively. And the associated strong coupling constant $\als$ are used. Here, the LO and NLO predictions are calculated with the LO and NLO PDF, respectively. The resummation predictions at  NLL and NNLL are with NLO PDF and NNLO PDF. It is because that the resummation terms include a bulk of the perturbative corrections appearing one order higher in $\als$ compared with the fixed order results.

\subsection{Determination of the scales}

\begin{figure}[t!]
  \includegraphics[width=0.49\textwidth]{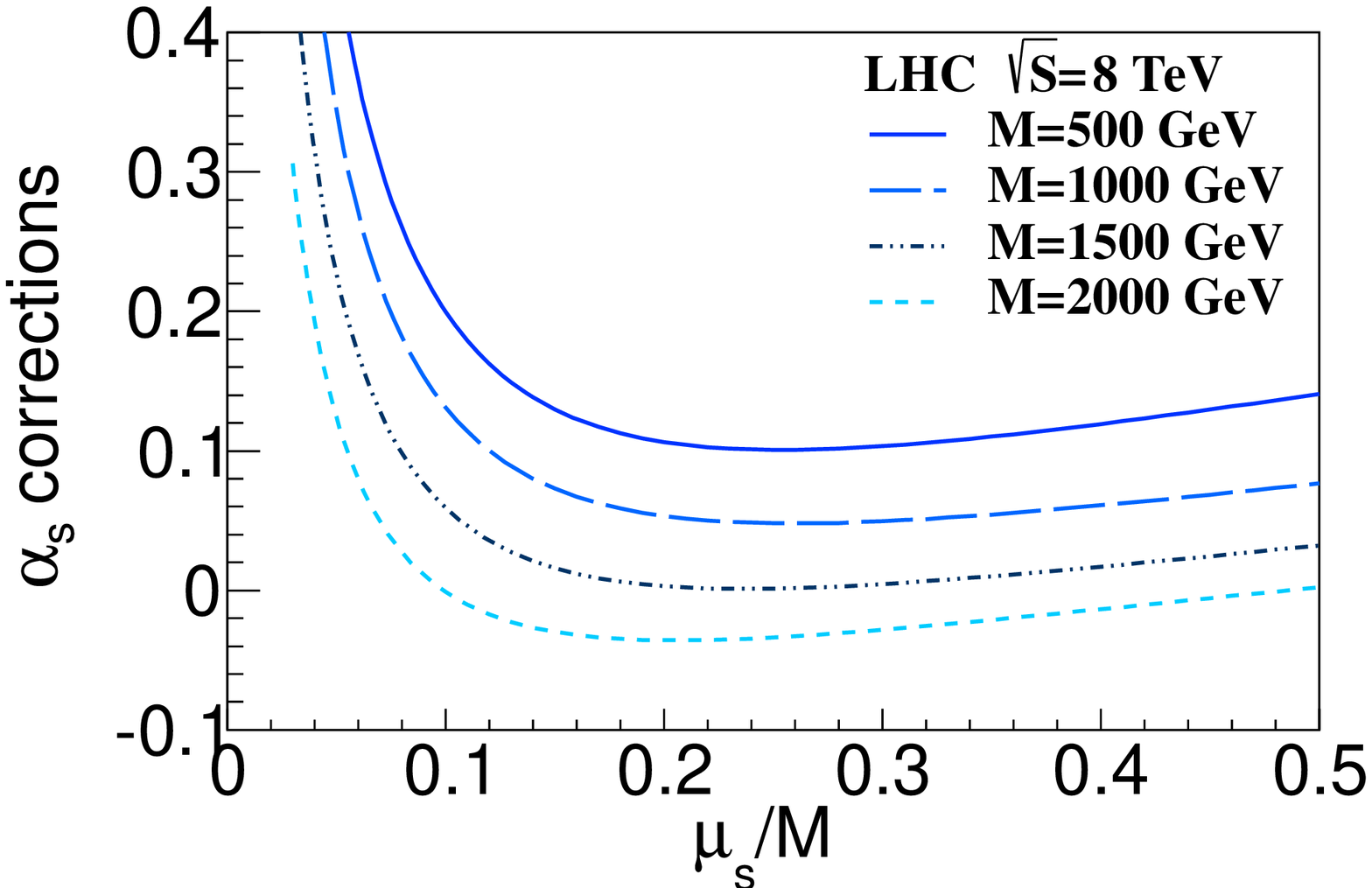}
  \includegraphics[width=0.49\textwidth]{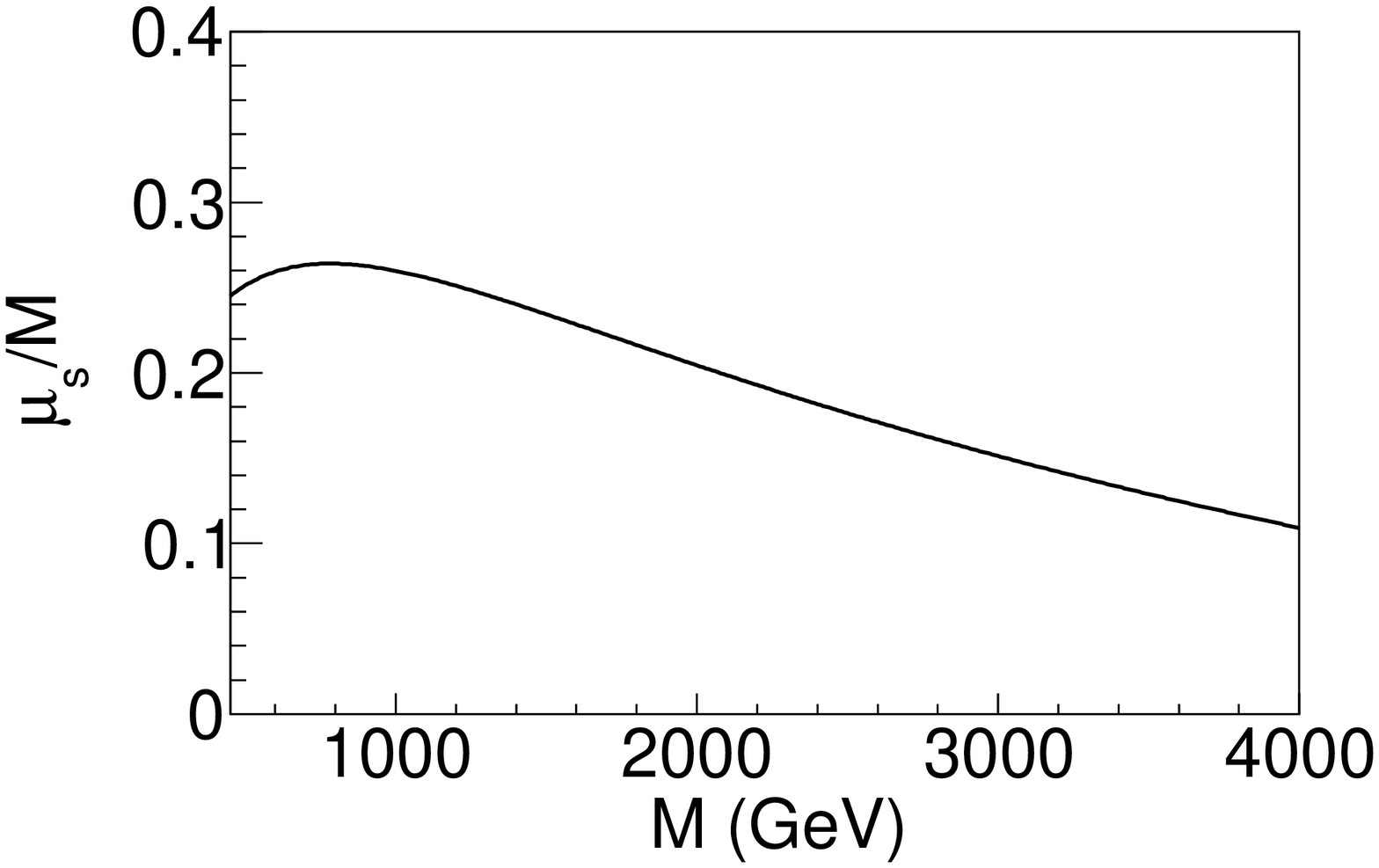}
  \includegraphics[width=0.49\textwidth]{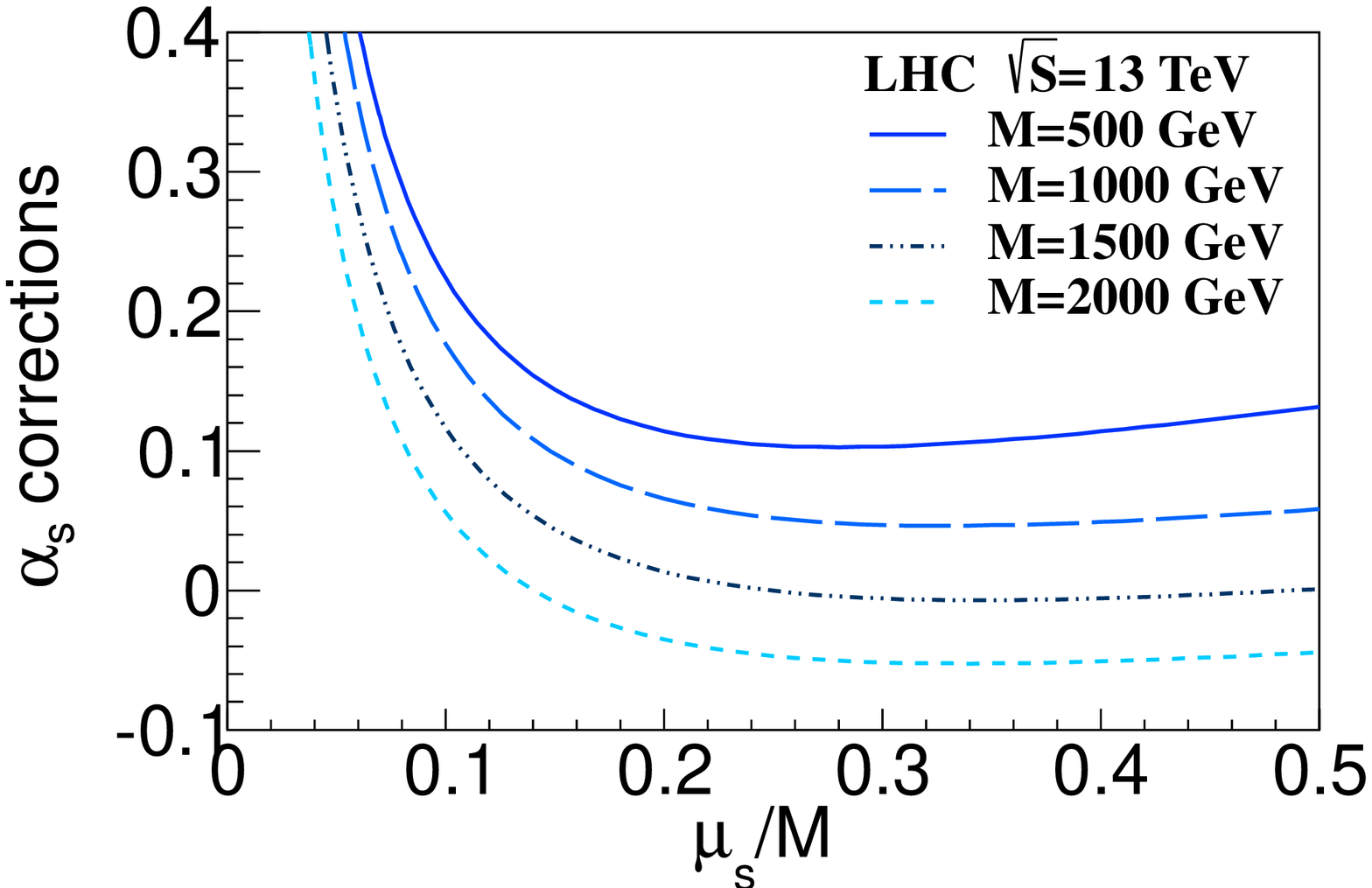}
  \includegraphics[width=0.49\textwidth]{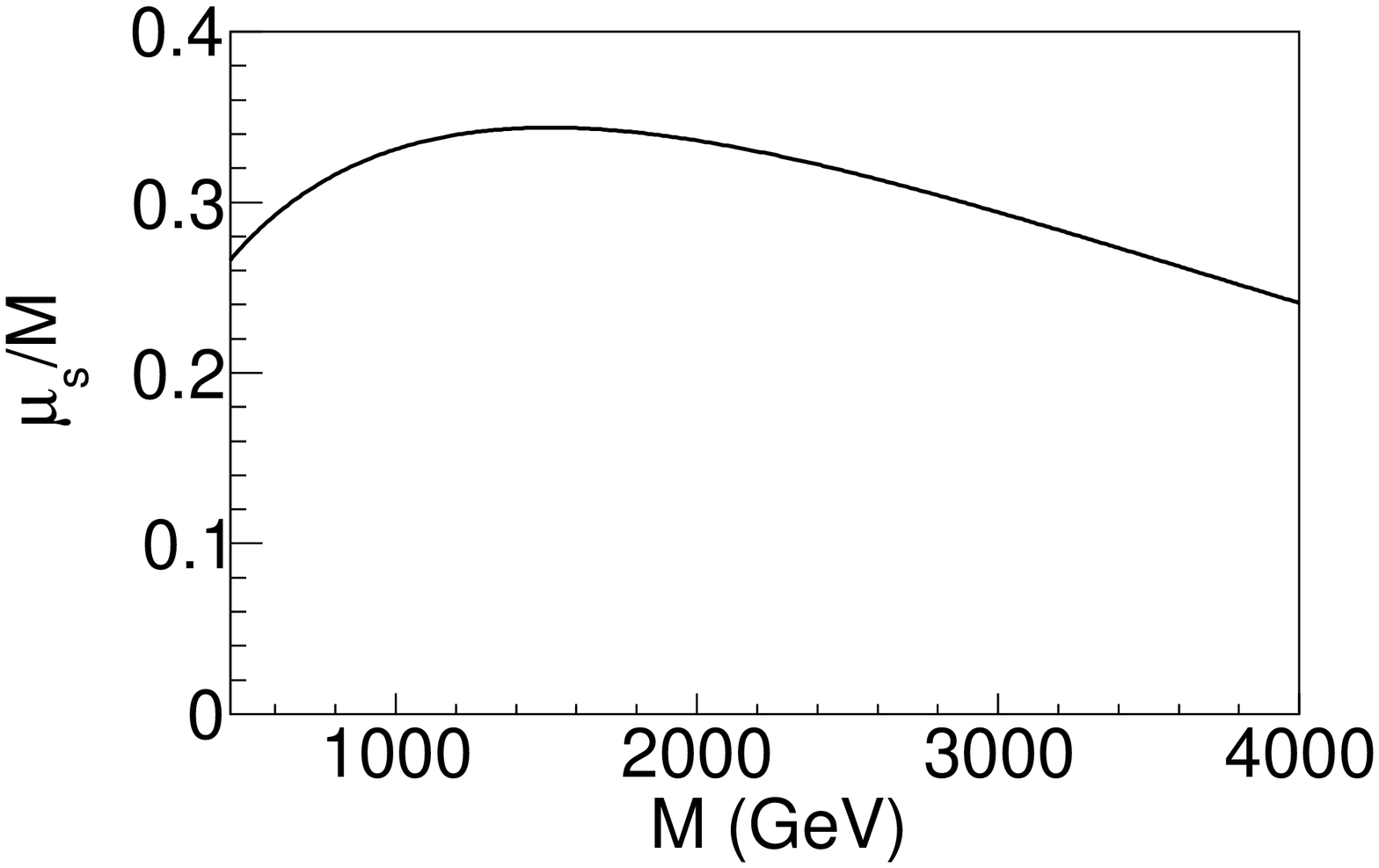}
 \vspace{-2ex}
  \caption{\label{fig:mus}  The left plots show the scale dependence of the one-loop soft function contribution divided by the LO cross section for $t\bar{t}W^+$ production with $M=500$, $1000$, $1500$ and $2000$ GeV. The right plots show the position of minimal corrections as a function of M.}
\end{figure}

The soft scale should be chosen where the perturbative series of the soft function are well behaved. To choose an appropriate soft scale, we pick out the contributions from the one-loop soft function in the NLO predictions, and then investigate these results divided by the LO cross sections. The scale dependences of these results are shown in the left hand of Fig.~\ref{fig:mus}, where the other scales are set as $\mu_h=M$ and $\mu_f=M$. The right figures in Fig.~\ref{fig:mus} give the position ($M/\mu_s$) of the minimum corrections as a function of M, where we choose as the default value of the soft scale. It can be found that the $\mu_s/M$ is between $0.26$ ($0.34$) and $0.11$ ($0.20$) at the 8(13) TeV LHC. The above default values of soft scale can be fitted by the formula
\begin{align}
  \label{eq:soft_scale}
  \mu_s^{\rm{def}} = M (1-\tau) b \tau^f \frac{(1+a\tau^{1/2})}{(1+c\tau^{1/2})^d} \ .
\end{align}
The fitting results are  $a=2.71$, $b=4.99$, $c=5.62$, $d=2.85$, $f=0.41$ at the 8 and 7 TeV LHC,  and $a=-0.50$, $b=1.65$, $c=0.49$, $d=7.66$, $f=0.25$ at the 13 and 14 TeV LHC.

\begin{figure}[t!]
  \includegraphics[width=0.49\textwidth]{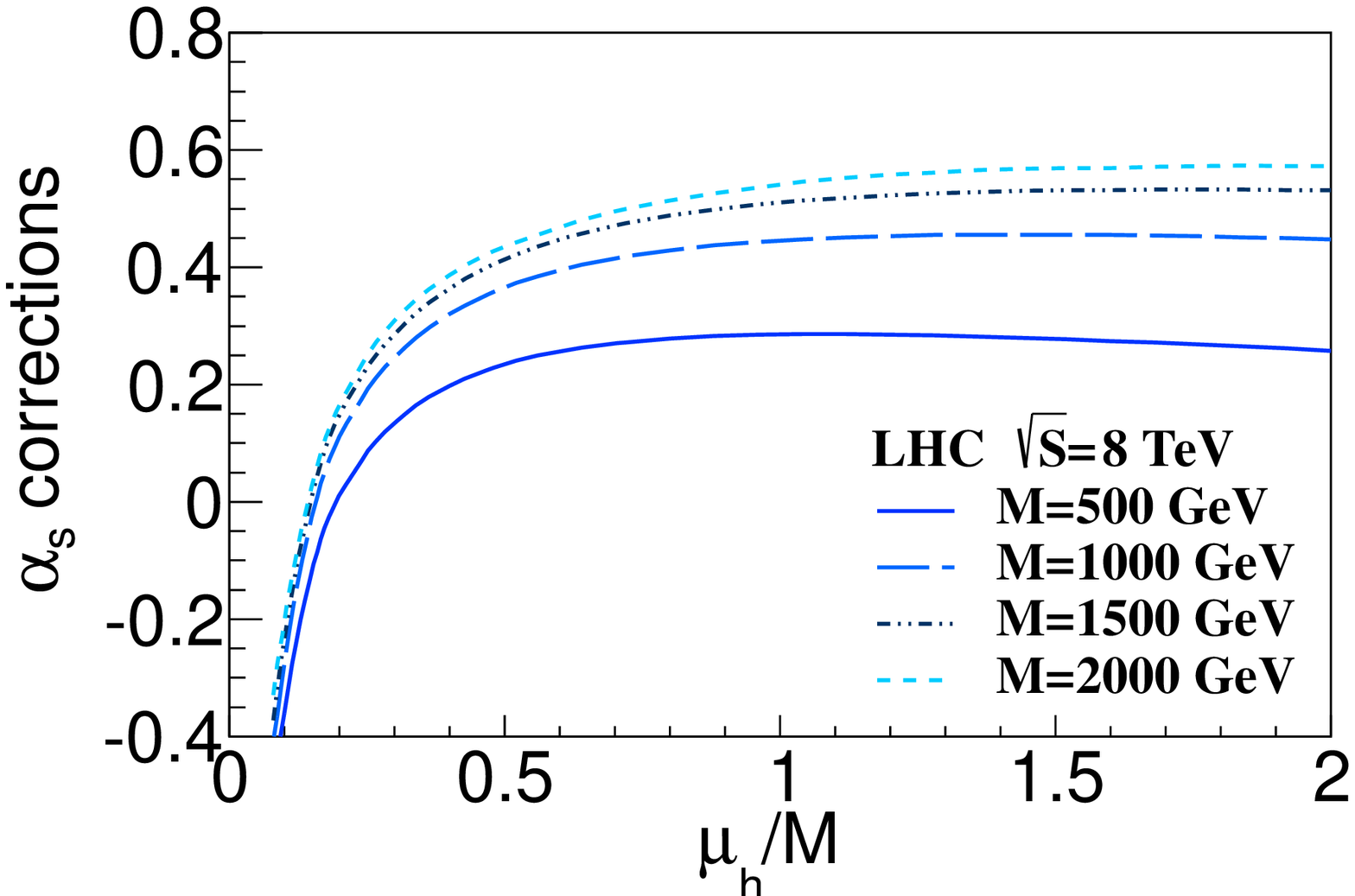}
  \includegraphics[width=0.49\textwidth]{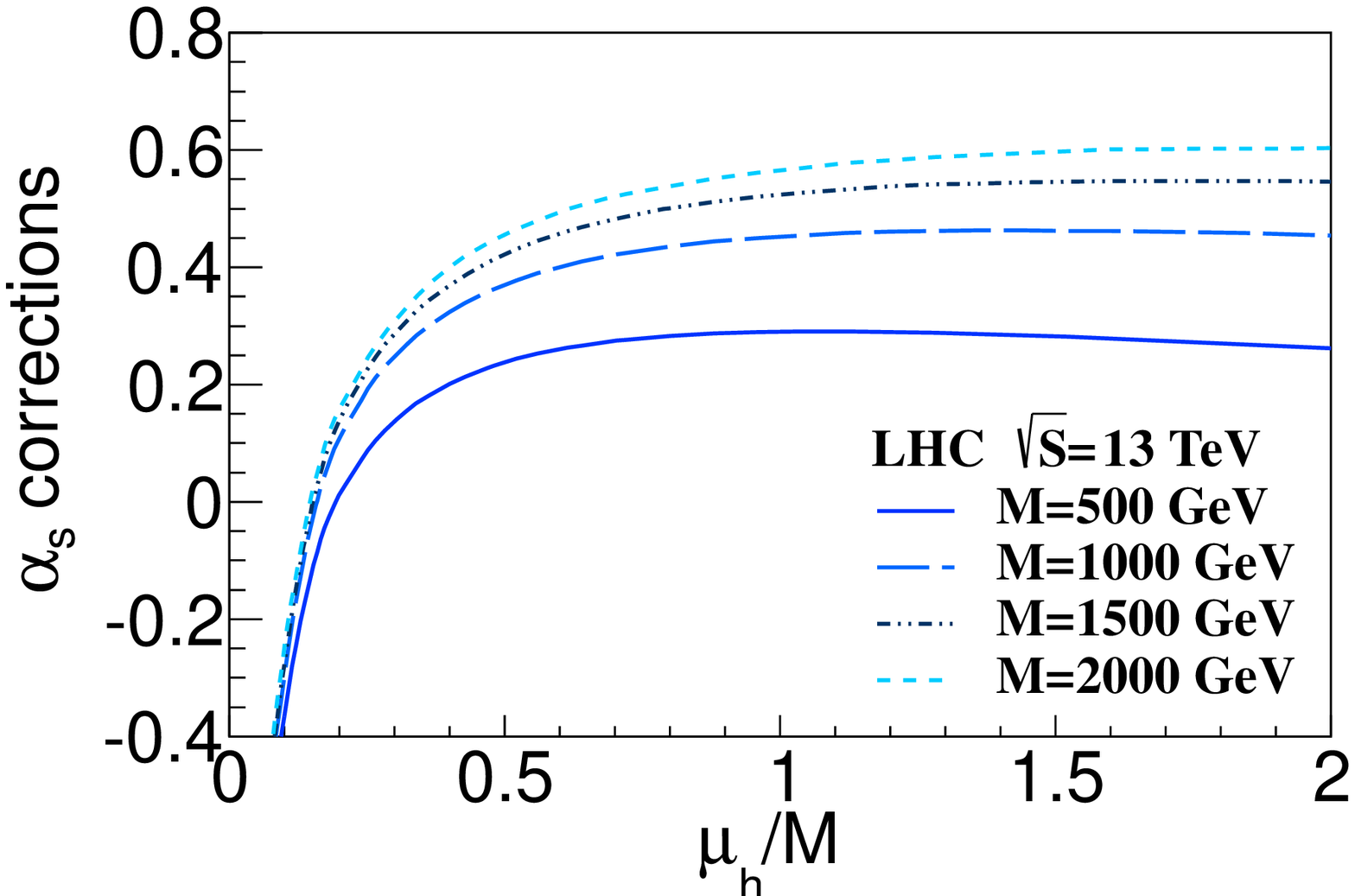}
 \vspace{-2ex}
  \caption{ \label{fig:muh} The scale dependence of  the one-loop hard function contribution of  NNLL predictions divided by the NLL cross section for $t\bar{t}W^+$ production  with $M=500$ GeV, $1000$ GeV, $1500$ GeV and $2000$ GeV at the 8 TeV (left) and 13 TeV (right) LHC. }
\end{figure}

Because there are three massive particles in the final states, the hard function depends on  several different scales, which leads that the most suitable hard scale is not apparent. Adopting the method in Ref.~\cite{Ahrens:2010zv}, we choose the stable hard scale  for hard function by looking at the corrections from the hard Wilson coefficient at different $\mu_h$. Figure~\ref{fig:muh} shows the scale dependence of the NNLL cross section which is defined as the NLO hard function contribution divided by the NLL results at the different $\mu_h$ with the choice of $\mu_f=M$ and $\mu_s$ determined by Eq.~(\ref{eq:soft_scale}). From Fig.~\ref{fig:muh}, we find that the corrections strongly depend on the hard scale and even become negative for small value of $\mu_h$, while it is more stable in the range $M/2 < \mu_h < 2 M$ at both the 8 and 13 TeV LHC. Therefore, we choose the default value of the hard scale as $\mu_h=M$. The same hard scale can also be obtained through the procedure for choosing the soft scales.

\begin{figure}[t!]
  \includegraphics[width=0.49\textwidth]{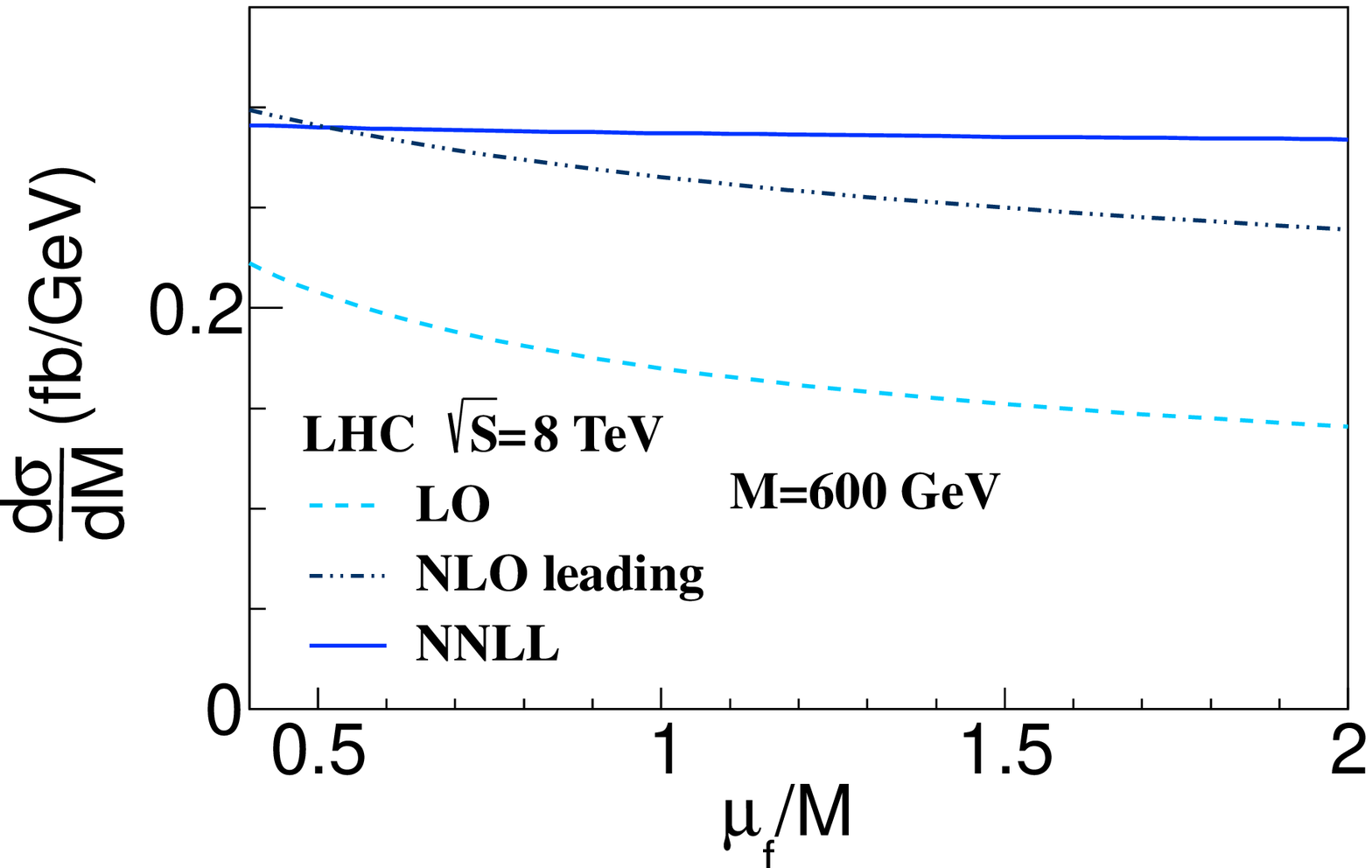}
  \includegraphics[width=0.49\textwidth]{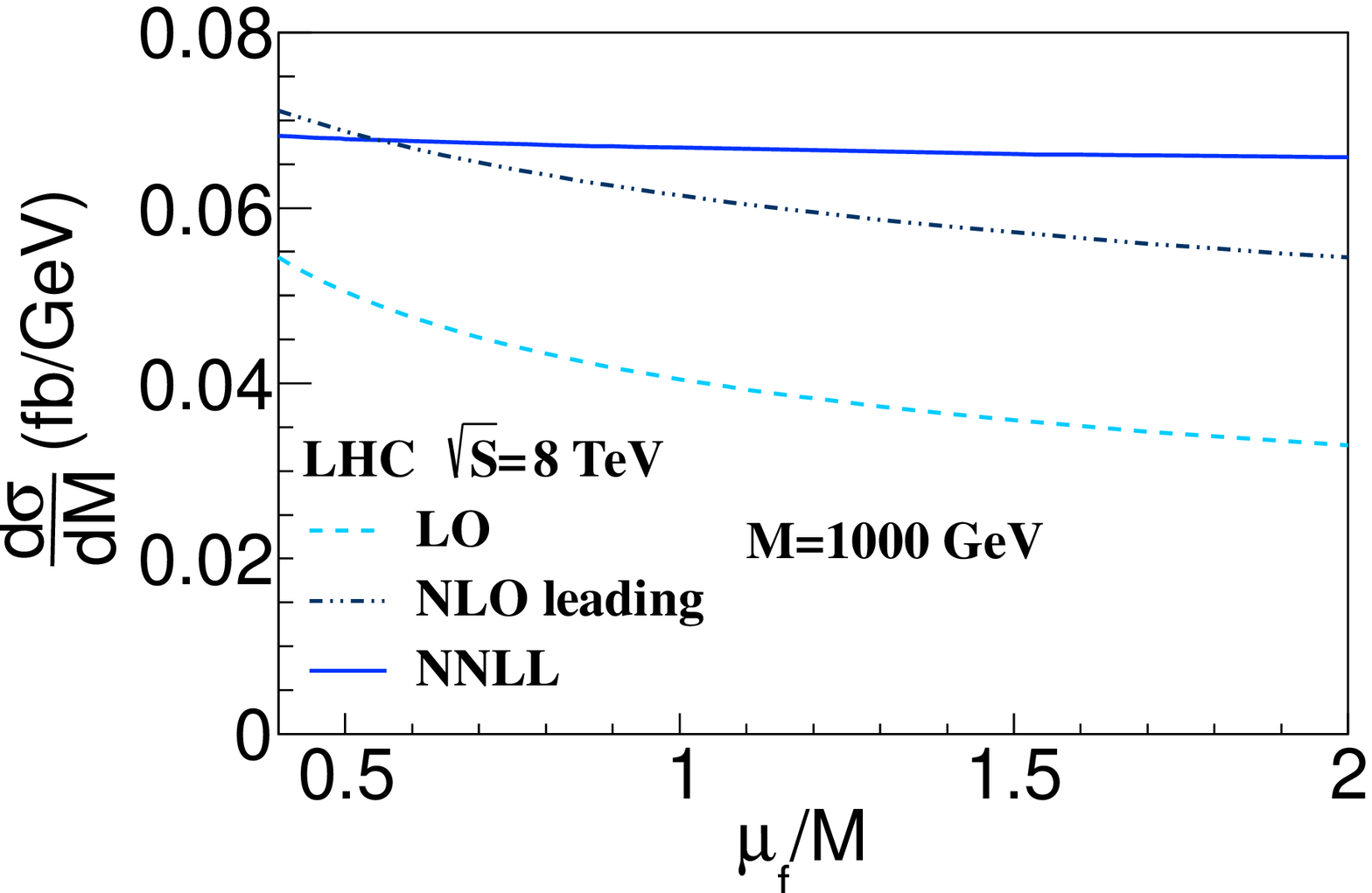}
  \includegraphics[width=0.49\textwidth]{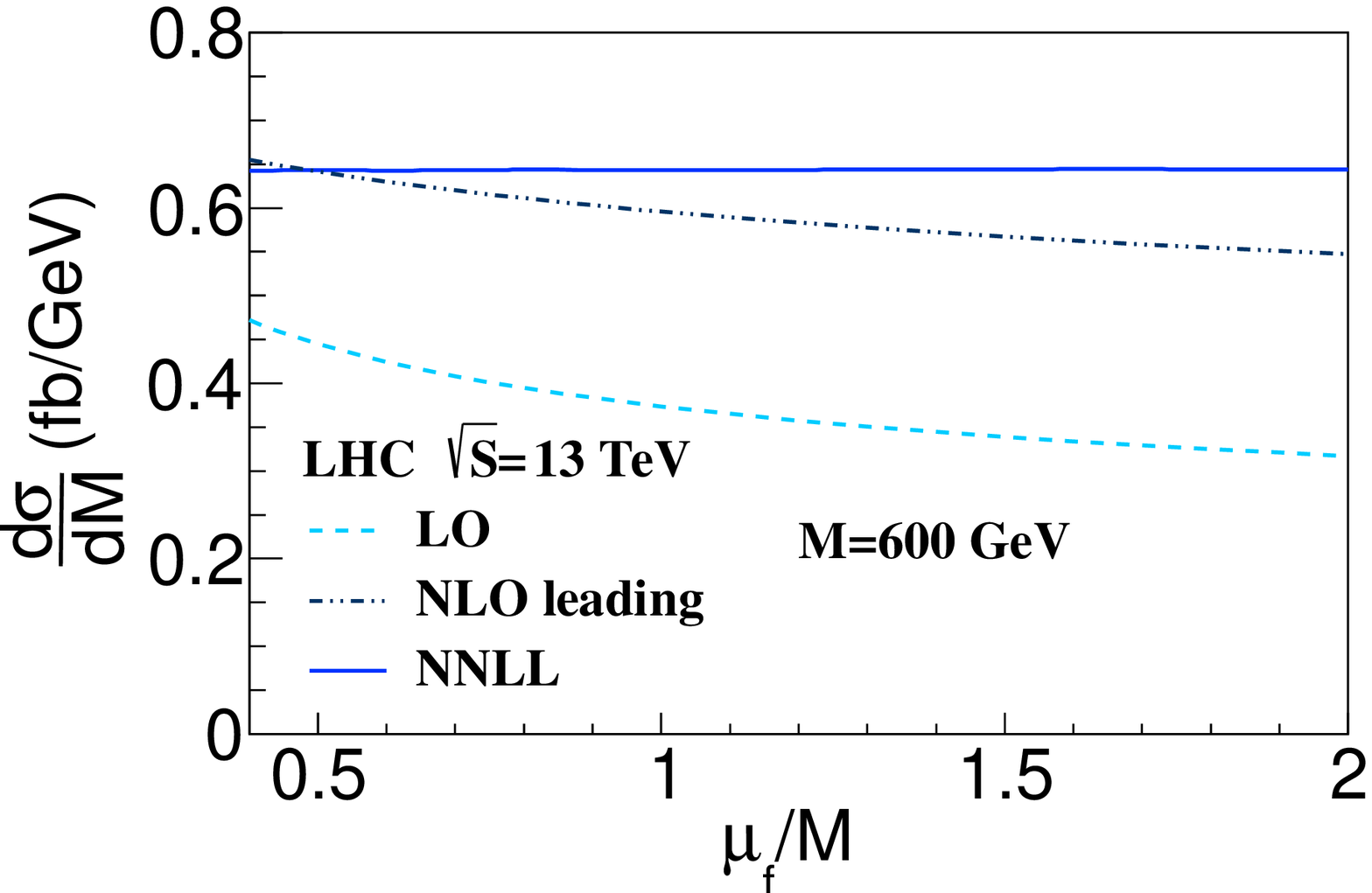}
  \includegraphics[width=0.49\textwidth]{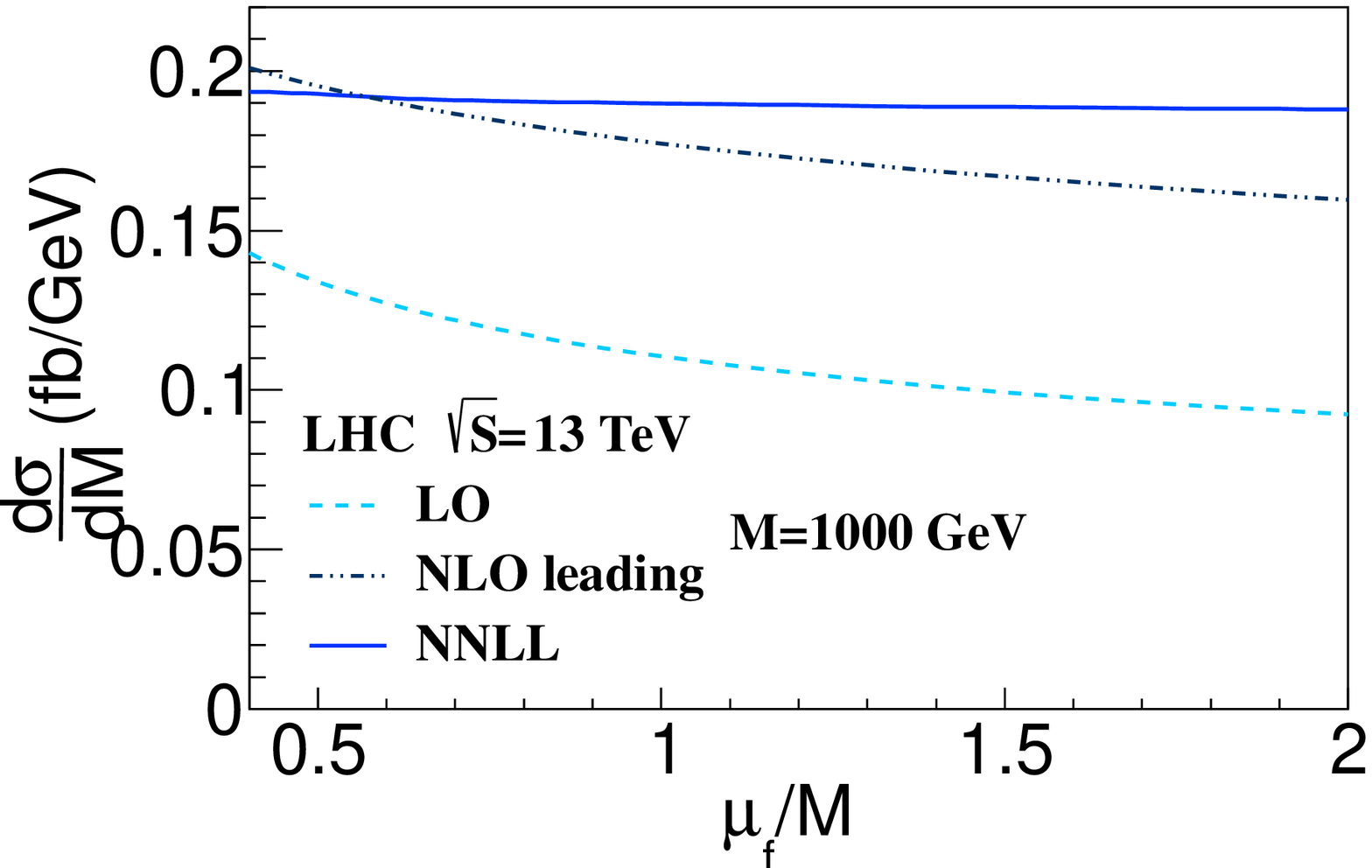}
 \vspace{-2ex}
  \caption{ \label{fig:muf} The factorization scale dependence of $d\sigma/dM$ at M=600 and M=1000GeV for $t\bar{t}W^+$ production. The solid line stands for the NNLL results. And the dashed and dash-dotted lines present the LO and NLO leading predictions, respectively.}
\end{figure}

Now we turn to the choice of factorization scale. As shown in Eq.~(\ref{eq:match}), both the resummed and fix order predictions depends on $\mu_f$.  To choose an appropriate factorization scale, we calculate the differential cross section $d\sigma/dM$  at M=600 and M=1000 GeV  as a function of $\mu_f$ for both fixed order  and resummation predictions, respectively, which are shown in Fig.~\ref{fig:muf}.  The leading singular terms are dominant at the NLO and have the similar dependence on $\mu_f$ with the NLO results. Therefore, we only show the $\mu_f$ dependence of the leading singular terms. It can be found that the LO results and the leading singular  terms are more unstable at small $\mu_f$, and the NNLL resummation results are almost independent on  $\mu_f$. We choose the $\mu_f=M$ as the default value for the factorization scale.

\subsection{RG improved predictions}

\begin{table}[t!]
  \centering
  \begin{tabular}{c c| c c c }
    \hline \hline
    &  & $\mu_h$ & $\mu_s$ & $\mu_f$
   \\
    \hline
    \multirow{2}{*} {$t\bar{t}W^+$} &
      8 TeV & $116.10^{+2\%}_{-0\%}$ & $ 116.10^{+1\%}_{-4\%}$ & $116.10^{+1\%}_{-1\%}$
     \\
      & 13 TeV & $293.53^{+2\%}_{-0\%}$ & $293.53^{+1\%}_{-4\%}$ & $293.53^{+0\%}_{-0\%}$
     \\
         \hline
    \multirow{2}{*} {$t\bar{t}W^-$} &
      8 TeV & $51.03^{+2\%}_{-0\%}$ & $ 51.03^{+2\%}_{-4\%}$ & $51.03^{+0\%}_{-0\%}$
     \\
      & 13 TeV & $150.23^{+2\%}_{-0\%}$ & $150.23^{+2\%}_{-4\%}$ & $150.23^{+1\%}_{-1\%}$
     \\
    \hline\hline
  \end{tabular}
  \caption{\label{tab:NNLL_scale} The $\mu_h$, $\mu_s$ and $\mu_f$ dependence of the NNLL total cross section.}
\end{table}
\begin{figure}[t!]
  \includegraphics[width=0.80\textwidth]{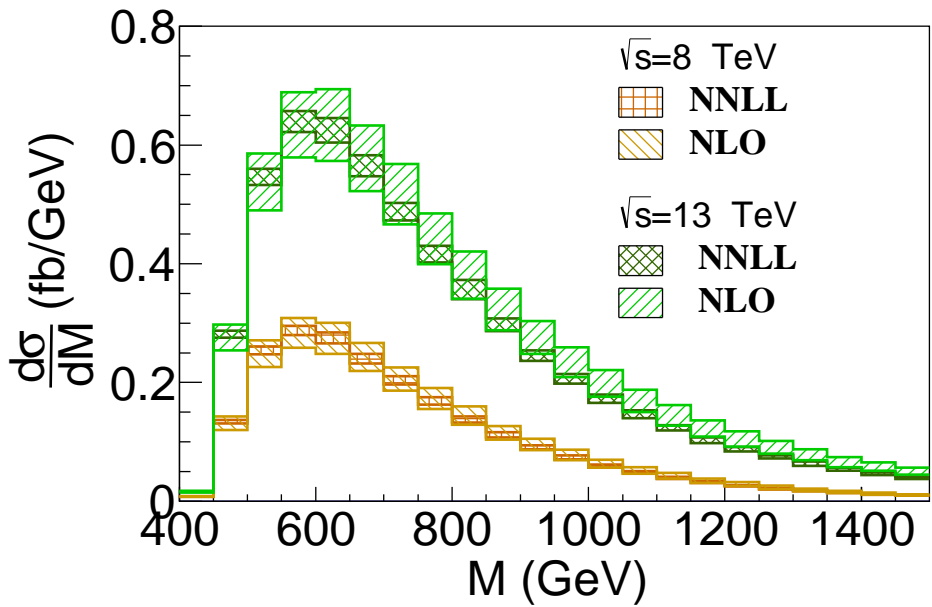}
 \vspace{-2ex}
  \caption{\label{fig:nnll} The NNLL and NLO invariant mass distribution for $t\bar{t}W^+$ production at the 8  and 13 TeV LHC. The bands are obtained through independently varying the scales $\mu_h$, $\mu_s$ and $\mu_f$ by a factor of two.}
\end{figure}

In Table~\ref{tab:NNLL_scale}, we show the dependence of the resummation predictions for the total cross section on scales. Here, the uncertainties are obtained through varying the scales independently in the range of $M/2 < \mu_f < 2 M$, $M/2 < \mu_h < 2 M$ and $\mu_s^{\rm{def}}/2< \mu_s < 2 \mu_s^{\rm{def}}$, respectively. It can be seen that the uncertainties arising from the variation of $\mu_h$, $\mu_s$ and $\mu_f$ are about 2\%, 4\%, and 1\%, respectively. Figure~\ref{fig:nnll} shows the NNLL invariant mass distribution for $t\bar{t}W^+$ production and its scale uncertainties. The NLO bands are due to the change of the factorization scale and the renormalization scales independently by a factor of 2 and the NNLL uncertainties arise from varying the scales in the range mentioned above.  From Fig.~\ref{fig:nnll}, we can see that the NNLL predictions reduce the scale uncertainties to a level of about 4\%, which are much smaller than the ones of NLO results.

\begin{figure}[t!]
  \includegraphics[width=0.80\textwidth]{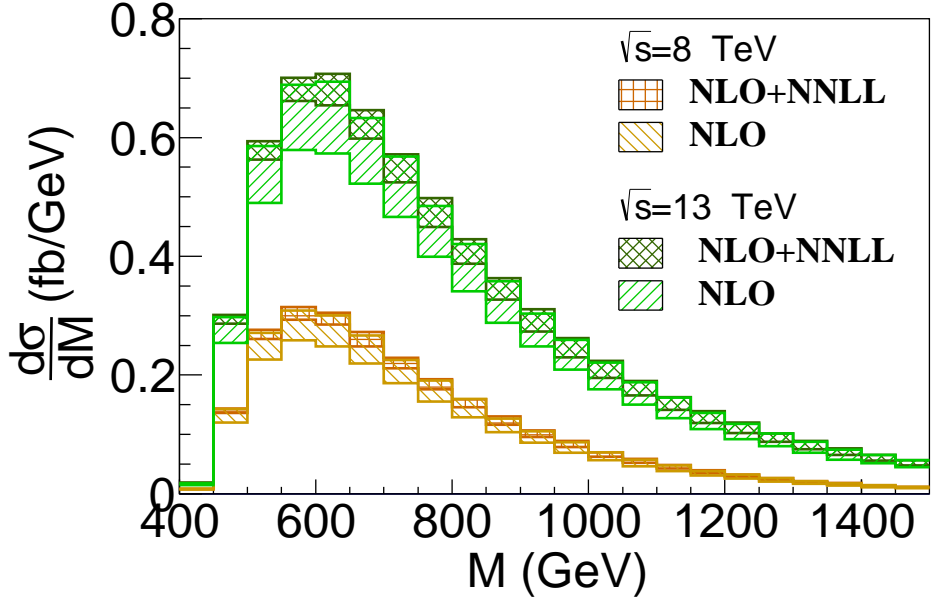}
 \vspace{-2ex}
  \caption{  \label{fig:nnll+nlo} The invariant mass distributions for the NNLL+NLO predictions for $t\bar{t}W^+$ production at 8 TeV and 13 TeV LHC. The bands stand  for the scale uncertainty. The bands represent the scale uncertainties. }
\end{figure}

\begin{table}[t!]
  \centering
  \begin{tabular}{c c| c c c c c c}
    \hline \hline
     & $\sqrt{s}$   &  LO (fb) & NLO (fb)  & NNLL (fb) &  NNLL$^{1}$(fb) & NNLL$^2$(fb) & K-factor \\
     \hline
     \multirow{4}{*}{$t\bar{t}W^+$}
     & 7 TeV & $52.62_{-15\%}^{+19\%}$ & $86.74_{-9\%}^{+10\%}$ & $86.57_{-4\%}^{+3\%}$ & $91.21_{-4\%}^{+3\%}$ & $93.02_{-5\%}^{+4\%}$ &  1.07 \\
     & 8 TeV & $69.71_{-15\%}^{+18\%}$ & $117.37_{-9\%}^{+10\%}$ & $116.10_{-4\%}^{+3\%}$ & $ 122.34_{-4\%}^{+2\%}$ &$125.91_{-4\%}^{+4\%}$ &  1.07  \\
     & 13 TeV & $171.40_{-14\%}^{+18\%}$ & $312.65_{-10\%}^{+11\%}$ & $ 293.52_{-4\%}^{+3\%}$ & $ 309.29_{-4\%}^{+3\%}$ &  $332.99_{-4\%}^{+5\%}$  & 1.07 \\
     & 14 TeV & $194.20_{-14\%}^{+18\%}$ & $358.38_{-10\%}^{+11\%}$ & $334.37_{-4\%}^{+2\%}$ &  $352.42_{-4\%}^{+2\%}$ & $382.55_{-5\%}^{+5\%}$ & 1.07 \\
    \hline
    \multirow{4}{*}{$t\bar{t}W^-$}
   & 7 TeV & $20.65_{-15\%}^{+19\%}$ & $35.60_{-10\%}^{+11\%}$ & $37.55_{-4\%}^{+3\%}$ & $39.51_{-4\%}^{+3\%}$ & $40.18_{-5\%}^{+4\%}$  & 1.13 \\
     & 8 TeV & $28.67_{-15\%}^{+19\%}$ & $50.29_{-10\%}^{+10\%}$ & $51.02_{-4\%}^{+2\%}$ & $ 53.79_{-4\%}^{+2\%}$ & $56.23_{-4\%}^{+4\%}$ &  1.10  \\
     & 13 TeV & $82.09_{-14\%}^{+18\%}$ & $154.28_{-10\%}^{+10\%}$  & $ 150.23_{-4\%}^{+3\%}$ & $ 158.13_{-4\%}^{+3\%}$ &  $169.86_{-4\%}^{+5\%}$  & 1.10\\
     & 14 TeV & $94.92_{-14\%}^{+18\%}$ & $180.46_{-10\%}^{+11\%}$ & $170.10_{-4\%}^{+2\%}$ &  $179.35_{-4\%}^{+3\%}$ & $194.62_{-5\%}^{+5\%}$  &  1.08\\
    \hline \hline
  \end{tabular}
  \caption{\label{tab:resum_ttw} The fixed order and resummed cross section for $t\bar{t} W^+$ and $t\bar{t}W^-$ production. The NNLL$^1$ and NNLL$^2$ represent the $q\bar{q}$ channel and total predictions at NLO+NNLL, respectively. The K-factor is defined as the NNLL$^2$ total cross section divided by the NLO one.}
\end{table}

Now, we turn to the NLO+NNLL predictions for invariant mass distribution as shown in Fig.~\ref{fig:nnll+nlo}.
For the fixed order calculation, we choose the default values for the renormalization scale and factorization scale as $\mu_r=\mu_f=M$ and vary the scales independent by a factor of 2. Note that the fixed order results is different with those in Refs.~\cite{Maltoni:2014zpa,Garzelli:2012bn} because of the different choice of scales.
It can be seen that the scale uncertainties of the NLO+NNLL distributions are reduced significantly. In the small invariant mass region for $M < 700$ GeV, the uncertainties are less than 1/3 of the ones of NLO results, while  in the large invariant mass region they are comparable. The uncertainties in the large invariant mass region mainly come from the subleading terms of the fixed order results in $qg$ channel. And the uncertainties due to the power suppressed corrections are about 1\% at the 7 TeV and 8 TeV LHC and less than 2\% at the 13 TeV and 14 TeV LHC.
 In Table~\ref{tab:resum_ttw}, we show the total cross sections for the fixed order and resummation predictions, where NNLL$^1$ and NNLL$^2$ mean the NNLL resummed predictions matched with the NLO results of $q\bar{q}$ channel and  the total NLO results, respectively.  The difference between NNLL$^1$ and NNLL$^2$ arises from the real corrections in $qg$ channel, which enlarges the scale uncertainties of the total cross section as mentioned before. As shown in Table~\ref{tab:resum_ttw}, comparing to the NLO results, the threshold resummation reduces the scale dependence from about $\pm 10 \%$ to about $\pm 5\%$ for both the $t\bar{t}W^+$ and $t\bar{t}W^-$ productions at different collider energies. The K-factor is defined as the NNLL$^2$ total cross sections divided by the NLO ones. Table~\ref{tab:resum_ttw} shows that the resummation increases the total cross section by about 7\%  for $t\bar{t}W^+$ production and about 10\% for $t\bar{t}W^-$ production.

\section{Conclusion}
\label{sec:conclude}
We have investigated the threshold resummation for  $t\bar{t}W^\pm$ production at the LHC with SCET. We briefly show the factorization formula in the threshold limit where the cross section can be factorized into a convolution of hard and soft functions and the PDFs. We present the analytical expression of the NLO soft function for this process, which can also be used in other processes of heavy quark pair  production in association with colorless particle, and we perform the threshold resummation  calculations by evolving the hard and soft functions to a common scale. Compared with NLO QCD results, the NLO+NNLL predictions increase the total cross section by about $7\%-13\%$ and reduce the dependence of the total and differential cross section on the scales significantly, which makes our results more reliable than the fixed-order results.

\begin{acknowledgments}
We would like to thank Hua Sheng Shao, Jian Wang, Li Lin Yang and Hua Xing Zhu for useful discussions. This work is supported by the National Natural Science Foundation of China under Grants No.~11375013 and No.~11135003.
\end{acknowledgments}

\bibliography{references}

\end{document}